\documentclass[12pt]{amsart}
\usepackage{amsfonts,amssymb,float}
\RequirePackage{ifpdf}
\usepackage{amsmath}
\usepackage{color}
\usepackage{pstricks}
\usepackage{pstricks-add}
\def\appendix#1{
\addtocounter{section}{1} \setcounter{equation}{0}
\renewcommand{\thesection}{\Alph{section}}
\section*{Appendix \thesection\protect\indent\quad
#1}
}
\renewcommand{\theequation}{\thesection.\arabic{equation}}

\catcode`\@=11
\def\marginnote#1{}

\newcount\hour
\newcount\minute
\newtoks\amorpm
\hour=\time\divide\hour by60 \minute=\time{\multiply\hour by60
\global\advance\minute by-\hour}
\edef\standardtime{{\ifnum\hour<12 \global\amorpm={am}%
        \else\global\amorpm={pm}\advance\hour by-12 \fi
        \ifnum\hour=0 \hour=12 \fi
        \number\hour:\ifnum\minute<10 0\fi\number\minute\the\amorpm}}
\edef\militarytime{\number\hour:\ifnum\minute<100\fi\number\minute}

\newcommand{\tcr}{\textcolor{red}}
\newcommand{\tcb}{\textcolor{blue}}

\def\draftlabel#1{{\@bsphack\if@filesw {\let\thepage\relax
      \xdef\@gtempa{\write\@auxout{\string
          \newlabel{#1}{{\@currentlabel}{\thepage}}}}}\@gtempa \if@nobreak
    \ifvmode\nobreak\fi\fi\fi\@esphack} \gdef\@eqnlabel{#1}}
    \def\@eqnlabel{}
\def\@vacuum{}
\def\draftmarginnote#1{\marginpar{\raggedright\scriptsize\tt#1}}

\def\draft{
  \oddsidemargin -.5truein
  \def\@oddfoot{\footnotesize \sl preliminary draft \hfil
    \rm\thepage\hfil\sl\today\quad\militarytime}
  \let\@evenfoot\@oddfoot \overfullrule 3pt
    \let\label=\draftlabel
    \let\marginnote=\draftmarginnote
  \def\@eqnnum{(\theequation)\rlap{\kern\marginparsep\tt\@eqnlabel}%
    \global\let\@eqnlabel\@vacuum}

  }

\voffset= - 0.5in \hoffset= - 1in

\def\be{\begin{equation}}
\def\ee{\end{equation}}
\def\bea{\begin{eqnarray}}
\def\eea{\end{eqnarray}}

\def\ocomma{{\phantom{\Bigm|}^{\phantom {X}}_{\raise-1.5pt\hbox{,}}\!\!\!\!\!\!\otimes}}
\newcommand{\col}[1]{{\raise-2pt\hbox{\tiny$\bullet$}\hskip -4.5pt \raise4pt\hbox{\tiny$\bullet$}{{#1}} \raise-2pt\hbox{\tiny$\bullet$}\hskip -4.5pt \raise4pt\hbox{\tiny$\bullet$}}}

\newcommand{\sheet}[2]{{\stackrel{{#1}}{{#2}}}}

\newtheorem{theorem}{Theorem}[section]
\newtheorem{lemma}[theorem]{Lemma}

\theoremstyle{definition}
\newtheorem{definition}[theorem]{Definition}
\newtheorem{example}[theorem]{Example}
\newtheorem{remark}[theorem]{Remark}

\allowdisplaybreaks

\begin{document}

\title[Cluster variables for affine Lie--Poisson systems]
{Cluster variables for affine Lie--Poisson systems}
\author{Leonid O. Chekhov$^{\ast}$}\thanks{$^{\ast}$Steklov Mathematical
Institute, Moscow, Russia, and Michigan State University, East Lansing, USA. Email: chekhov@msu.edu.}

\begin{abstract}
We show that having any planar (cyclic or acyclic) directed network on a disc with the only condition that all $n_1+m$ sources are separated from all $n_2+m$ sinks, we can construct a cluster-algebra realization of elements of an affine Lie--Poisson algebra $R(\lambda,\mu)\sheet{1}T(\lambda)\sheet{2}T(\mu)=\sheet{2}T(\mu)\sheet{1}T(\lambda)R(\lambda,\mu)$ with $(n_1\times n_2)$-matrices $T(\lambda)$ corresponding to a planar directed network on an annulus. Upon satisfaction of some invertibility conditions, we can extend this construction to realizations of a quantum loop algebra. Having the quantum loop algebra we can also construct a realization of the twisted Yangian algebra, or that of the quantum reflection equation. Every such planar network therefore corresponds to a symplectic leaf  of the corresponding infinite-dimensional algebra. 
\end{abstract}

\dedicatory{Dedicated to the memory of Serezha Natanzon}

\maketitle

\section{Introduction and known facts}
In the recent paper \cite{ChSh2}, the author together with Michael Shapiro developed a quantum version of Poisson relations between elements of transport matrices for directed networks on a disc. One of results of this paper reads as follows.

Let $\mathcal N$ be a directed network on a disc with arbitrary distribution  of sources and sinks. Let $\omega_i$ be variables assigned to two-dimensional simplices (both inner and boundary) of the simplicial cell decomposition of the disc with the network $\mathcal N$. The variables $\omega_i$ then satisfy canonical commutation relations $q^{\varepsilon_{ij}}\omega_i \omega_j = q^{-\varepsilon_{ij}}\omega_j \omega_i$ with $q\in\mathbb C$ and with ${\varepsilon_{ij}}$ being elements of the exchange matrix of the corresponding quiver.  

Define the quantum transport matrix element $T_{a\to c}$ from the source labeled $a$ to the sink labeled $c$ as
$$
T_{a\to c}=\sum_{\text{all directed}\atop \text{paths $a\to c$}}(-1)^{\#\text{self-intersections}} \col{\prod_{\text{cells to the right}\atop \text{of the path}} \omega_i},
$$
where the sum is taken over all directed paths from $a$ to $c$ including those with repetitions and self-intersections (this is possible only if a network $\mathcal N$ contains closed directed cycles), in which case this sum can be infinite, and the product of $\omega_i$ over cells is then also taken with repetitions; the result is a rational expression for any finite network. Double dots around the product stay for the Weyl ordering of the corresponding product; recall that 
$$
\col{\omega_i \omega_j }:=q^{\varepsilon_{ij}}\omega_i \omega_j = q^{-\varepsilon_{ij}}\omega_j \omega_i=\col{\omega_j \omega_i},
$$
so variables always commute under the sign of the Weyl ordering. G. Schrader  and  A. Shapiro  showed~\cite{Gus-Al} that these quantum elements are stable under the quantum mapping class group action. 

\noindent
{\bf Theorem.} \cite{ChSh2}. {\it For any planar network $\mathcal N$ on a disc, the above defined elements of the transport matrix satisfy the following simple algebra: having two distinct sources $a$ and $b$ and two distinct sinks $c$ and $d$, for the (clockwise) cyclically ordered sources and sinks $a\to c\to b\to d\to a$ we have $T_{a\to c}T_{b\to d}=T_{b\to d}T_{a\to c}$ and $T_{a\to d}T_{b\to c}=T_{b\to c}T_{a\to d}$, in other words, all such elements commute, and for the cyclic ordering of sources and sinks $a\to b\to c\to d\to a$, we have that $T_{a\to c}T_{b\to d}=T_{b\to d}T_{a\to c}$, whereas the only nontrivial relation is between $T_{a\to d}$ and $T_{b\to c}$: 
$$
[T_{a\to d},T_{b\to c}]=(q-q^{-1})T_{a\to c}T_{b\to d}.
$$
If $a=b$, we have $T_{a\to d}T_{a\to c}=q T_{a\to c}T_{a\to d}$ and if $c=d$, we have $T_{a\to c}T_{b\to c}=q T_{b\to c}T_{a\to c}$.}

If we impose the restriction that all $m_1$ sources $a_j$ are separated from all $m_2$ sinks $c_i$ (so they are located on two different ``halves''  of the disc boundary, which, in particular, means that the first case of two commuting transport matrix elements in the above theorem is never realized) then the commutation relations between elements can be conveniently written in the $R$-matrix form: for $\mathcal M$ the $(m_2\times m_1)$-matrix with entries $\mathcal M_{i,j}:=T_{a_j\to c_i}$, we can present the commutation relations in the Lie--Poisson form
\be
\label{RMM}
R_{n_2} \sheet{1}{\mathcal M}\sheet{2}{\mathcal M}=\sheet{2}{\mathcal M}\sheet{1}{\mathcal M}R_{n_1},
\ee
where we use the standard notation for the direct product of two transport matrices sharing the same quantum space and $R_k$ is the standard Kulish--Sklyanin  trigonometric $R$-matrix
\be\label{R-matrix1}
R_k= \sum_{1\le i,j\le k} \sheet{1}e_{ii}\otimes \sheet{2} e_{jj}+(q-1)\sum_{1\le i\le k} \sheet{1}e_{ii}\otimes \sheet{2} e_{ii} +(q{-}q^{-1})\sum_{1\le j<i \le k} \sheet{1}e_{ij}\otimes \sheet{2} e_{ji}
\ee
satisfying the Yang--Baxter equation and determining an automorphism of the direct product $V_k\otimes V_k$ of two $k$-dimensional vector spaces. 

\begin{figure}[h]
	\begin{pspicture}(-3,-3)(4,4){
		\newcommand{\LEFTDOWNARROW}{%
			{\psset{unit=1}
				\rput(0,0){\psline[doubleline=true,linewidth=1pt, doublesep=1pt, linecolor=black]{<-}(0,0)(.765,.45)}
		}}
		\newcommand{\DOWNARROW}{%
	{\psset{unit=1}
					\rput(0,0){\psline[doubleline=true,linewidth=1pt, doublesep=1pt, linecolor=black]{->}(0,0.1)(0,-0.566)}
		\put(0,0){\pscircle[fillstyle=solid,fillcolor=lightgray]{.1}}
}}
		\newcommand{\LEFTUPARROW}{%
	{\psset{unit=1}
		\rput(0,0){\psline[doubleline=true,linewidth=1pt, doublesep=1pt, linecolor=black]{->}(0,0)(-.765,.45)}
}}
	\newcommand{\STARUP}{
			{\psset{unit=1}
	\rput(0,0){\psline[doubleline=true,linewidth=1pt, doublesep=1pt, linecolor=black]{<-}(0.06,-0.03)(.4,-.24)}
	\rput(0,0){\psline[doubleline=true,linewidth=1pt, doublesep=1pt, linecolor=black]{<-}(0,0.1)(0,.466)}
	\rput(0,0){\psline[doubleline=true,linewidth=1pt, doublesep=1pt, linecolor=black]{->}(0,0)(-.5,-.26)}
	\put(0,0){\pscircle[fillstyle=solid,fillcolor=black]{.1}}
	\put(0,.566){\pscircle[fillstyle=solid,fillcolor=lightgray]{.1}}
}}
		\newcommand{\PATGEN}{%
			{\psset{unit=1}
				\rput(0,0){\psline[linecolor=blue,linewidth=2pt]{->}(0,0)(.45,.765)}
				\rput(0,0){\psline[linecolor=blue,linewidth=2pt]{->}(1,0)(0.1,0)}
				\rput(0,0){\psline[linecolor=blue,linewidth=2pt]{->}(0,0)(.45,-.765)}
				\put(0,0){\pscircle[fillstyle=solid,fillcolor=red]{.1}}
		}}
		\newcommand{\PATLEFT}{%
			{\psset{unit=1}
				\rput(0,0){\psline[linecolor=blue,linewidth=2pt,linestyle=dashed]{->}(0,0)(.45,.765)}
				\rput(0,0){\psline[linecolor=blue,linewidth=2pt]{->}(1,0)(0.1,0)}
				\rput(0,0){\psline[linecolor=blue,linewidth=2pt]{->}(0,0)(.45,-.765)}
				\put(0,0){\pscircle[fillstyle=solid,fillcolor=red]{.1}}
		}}
		\newcommand{\PATRIGHT}{%
			{\psset{unit=1}
				\rput(0,0){\psline[linecolor=blue,linewidth=2pt,linestyle=dashed]{->}(0,0)(.45,-.765)}
				\put(0,0){\pscircle[fillstyle=solid,fillcolor=red]{.1}}
		}}
		\newcommand{\PATBOTTOM}{%
			{\psset{unit=1}
				\rput(0,0){\psline[linecolor=blue,linewidth=2pt]{->}(0,0)(.45,.765)}
				\rput(0,0){\psline[linecolor=blue,linewidth=2pt,linestyle=dashed]{->}(1,0)(0.1,0)}
				\put(0,0){\pscircle[fillstyle=solid,fillcolor=red]{.1}}
		}}
		\newcommand{\PATTOP}{%
			{\psset{unit=1}
				\rput(0,0){\psline[linecolor=blue,linewidth=2pt]{->}(1,0)(0.1,0)}
				\rput(0,0){\psline[linecolor=blue,linewidth=2pt]{->}(0,0)(.45,-.765)}
				\put(0,0){\pscircle[fillstyle=solid,fillcolor=red]{.1}}
		}}
		\newcommand{\PATBOTRIGHT}{%
			{\psset{unit=1}
				\rput(0,0){\psline[linecolor=blue,linewidth=2pt]{->}(0,0)(.45,.765)}
				\put(0,0){\pscircle[fillstyle=solid,fillcolor=red]{.1}}
				\put(.5,0.85){\pscircle[fillstyle=solid,fillcolor=red]{.1}}
		}}
		\newcommand{\PATNORTH}{%
			{\psset{unit=1}
				\rput(0,0){\psline[linecolor=blue,linewidth=2pt,linestyle=dashed]{->}(-.45,-.765)(0,0)}
				\rput(0,0){\psline[linecolor=blue,linewidth=2pt,linestyle=dashed]{<-}(.45,-.765)(0,0)}
				\put(0,0){\pscircle[fillstyle=solid,fillcolor=red]{.1}}
		}}
		\newcommand{\PATSW}{%
			{\psset{unit=1}
				\rput(0,0){\psline[linecolor=blue,linewidth=2pt,linestyle=dashed]{->}(0,0)(.45,.765)}
				\rput(0,0){\psline[linecolor=blue,linewidth=2pt,linestyle=dashed]{<-}(0.1,0)(0.9,0)}
				\put(0,0){\pscircle[fillstyle=solid,fillcolor=red]{.1}}
		}}
		\newcommand{\PATSE}{%
			{\psset{unit=1}
				\rput(0,0){\psline[linecolor=blue,linewidth=2pt,linestyle=dashed]{<-}(0,0)(-.45,.765)}
				\rput(0,0){\psline[linecolor=blue,linewidth=2pt,linestyle=dashed]{<-}(-0.9,0)(-0.1,0)}
				\put(0,0){\pscircle[fillstyle=solid,fillcolor=red]{.1}}
		}}
		\multiput(-2.5,-0.85)(0.5,0.85){4}{\PATLEFT}
		\multiput(-2,-1.7)(1,0){4}{\PATBOTTOM}
		\put(-0.5,2.55){\PATTOP}
		\put(0,3.4){\PATNORTH}
		\multiput(-1.5,-0.85)(1,0){4}{\PATGEN}
		\multiput(-1,0)(1,0){3}{\PATGEN}
		\multiput(-.5,0.85)(1,0){2}{\PATGEN}
		\put(0,1.7){\PATGEN}
		\multiput(-1.5,-0.85)(1,0){4}{\PATGEN}
		\multiput(0.5,2.55)(0.5,-0.85){4}{\PATRIGHT}
		\put(2,-1.7){\PATBOTRIGHT}
		\put(-3,-1.7){\PATSW}
		\put(3,-1.7){\PATSE}
		\multiput(-2,-1.176)(1.0,0){5}{\STARUP}
		\multiput(-1.5,-0.335)(1.0,0){4}{\STARUP}
		\multiput(-1.0,0.5)(1.0,0){3}{\STARUP}
		\multiput(-.5,1.4)(1.0,0){2}{\STARUP}
		\put(0,2.3){\STARUP}
		\multiput(2.6,-1.4)(-0.5,.85){6}{\LEFTDOWNARROW}
		\multiput(-2.6,-1.4)(0.5,.85){6}{\LEFTUPARROW}
		\multiput(-2.5,-1.5)(1.0,0){6}{\DOWNARROW}
		\put(1.2,3.2){\makebox(0,0)[br]{\hbox{{$1$}}}}
		\put(1.7,2.4){\makebox(0,0)[br]{\hbox{{$2$}}}}
		\put(2.2,1.6){\makebox(0,0)[br]{\hbox{{$3$}}}}
		\put(2.7,0.8){\makebox(0,0)[br]{\hbox{{$4$}}}}
		\put(3.2,-0.1){\makebox(0,0)[br]{\hbox{{$5$}}}}
		\put(3.7,-1.0){\makebox(0,0)[br]{\hbox{{$6$}}}}

		\put(-1.2,3.2){\makebox(0,0)[br]{\hbox{{$1'$}}}}
		\put(-1.7,2.4){\makebox(0,0)[br]{\hbox{{$2'$}}}}
		\put(-2.2,1.6){\makebox(0,0)[br]{\hbox{{$3'$}}}}
		\put(-2.7,0.8){\makebox(0,0)[br]{\hbox{{$4'$}}}}
		\put(-3.2,-0.1){\makebox(0,0)[br]{\hbox{{$5'$}}}}
		\put(-3.7,-1.0){\makebox(0,0)[br]{\hbox{{$6'$}}}}

		\put(-2.4,-2.6){\makebox(0,0)[br]{\hbox{{$1''$}}}}
		\put(-1.4,-2.6){\makebox(0,0)[br]{\hbox{{$2''$}}}}
		\put(-0.4,-2.6){\makebox(0,0)[br]{\hbox{{$3''$}}}}
		\put(0.6,-2.6){\makebox(0,0)[br]{\hbox{{$4''$}}}}
		\put(1.6,-2.6){\makebox(0,0)[br]{\hbox{{$5''$}}}}
		\put(2.6,-2.6){\makebox(0,0)[br]{\hbox{{$6''$}}}}
	}
	\end{pspicture}
	\caption{\small
		The network $N$ dual to the quiver of Fock--Goncharov parameters for $SL_6$ in the triangle $\Sigma_{0,1,3}$. Double arrows are edges of the directed network,  cluster variables correspond to faces of $N$ (vertices of the graph dual to $N$), and the dual graph is the corresponding quiver: its solid directed edges correspond to $\varepsilon_{ij}=1$ and its dashed directed edges---to $\varepsilon_{ij}=1/2$.
	}
	\label{fi:triangle}
\end{figure}
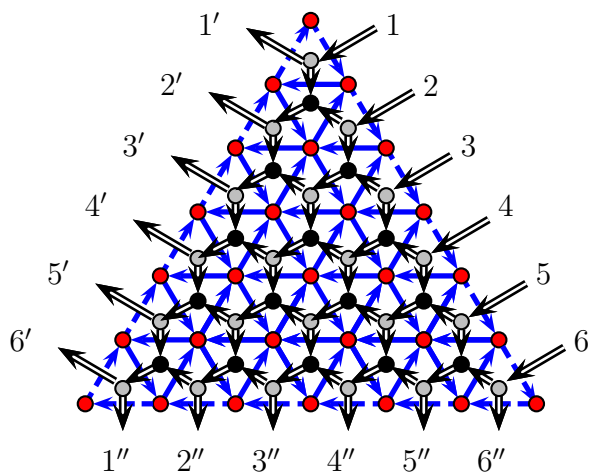

One of important corollaries of formula (\ref{RMM}) is the observation in \cite{ChSh2} that having $\mathcal M$ of the size $2n_2\times n_1$, we can split it into two blocks of equal sizes $n_2\times n_1$, the upper block $M_1$ and the lower block $M_2$. These blocks satisfy a closed algebra, and one of its important examples is the quantum algebra of monodromies of $SL_n$ Fuchsian system on the disc with three marked points on the boundary, (a triangle)  \cite{CMR}, or a disc with three complete flags in Fock--Goncharov description of higher Teichm\"uller spaces \cite{FG1} depicted in Fig.~\ref{fi:triangle}. For any such network, the matrix product $\mathbb A:=[M_1^{\text{T}}]M_2$, which is a square matrix of size $n_1\times n_1$, satisfies the {\em quantum reflection equation}
\be
\label{RARA}
R_{n_1}\sheet{1}{\mathbb A} R_{n_1}^{\text{t}_1}\sheet{2} {\mathbb A}=
\sheet{2}{\mathbb A} R_{n_1}^{\text{t}_1}\sheet{1} {\mathbb A}R_{n_1},
\ee
with $\text{t}_1$ indicating partial transposition w.r.t. the first space. For transport matrices corresponding to the complete triangular network with sides of length $n_1=n_2$, $n_1$ sources, and $2n_1$ sinks depicted in Fig.~\ref{fi:triangle}, the above matrix $\mathbb A$ is upper-triangular, and its entries can be identified with {\em quantum geodesic functions} $G_{i,j}$, while the corresponding Gavrilik--Klimyk--Nelson--Regge--Ugaglia algebras \cite{GK91}, \cite{NR}, \cite{NRZ}, \cite{Ugaglia} as they are known in mathematical physics, follow from the quantum skein relations for quantum geodesic functions. Another description of the same object based on the groupoid of upper-triangular matrices was proposed by Bondal \cite{Bondal}, who also classified all symplectic leaves of this algebra using the Jordan form of the matrix combination ${\mathbb A}{\mathbb A}^{-\text{T}}$ (note that we are about to return to this construction in the forthcoming paper \cite{ChShSh}). Some of these symplectic leaves were identified (see \cite{ChF2}, \cite{ChF3}) with systems of geodesic functions on Riemann surfaces of arbitrary genus and with one or two holes, but for $n_1\ge 6$ these geometric symplectic leaves have dimensions smaller than those of general symplectic leaves. The approach in \cite{ChSh2} is more general and provides Darboux coordinate parameterizations in terms of quantum cluster variables presumably for all symplectic leaves of the algebra (\ref{RARA}). 

A prominent generalization of relations (\ref{RMM}) is based on affinnization: $\mathcal M\mapsto T(\lambda)=\sum_{k=0}^\infty T_k \lambda^{-k}$ with $T_0=\mathcal M$, so we lift the situation to an infinite-dimensional space; the corresponding matrices must then satisfy the equation (\ref{RmuTT}) with the $R$-matrix $R(\lambda,\mu):= \lambda R^{-\text{T}} - \mu R$. An affine analogue of the quantum reflection equation can be constructed in a regular way out of quantum loop algebra (see, e.g., the monograph by Molev \cite{Molev}). 

Affinnization of Lie--Poisson algebras of transport matrices was constructed by Gekhtman, Shapiro, and Vainstein \cite{GSV},\cite{GSV2} who proved the semiclassical analogue of these algebras for a directed network on a cylinder (annulus) in \cite{GSV3}, and by Chekhov and Mazzocco \cite{ChM}, who embedded the algebra of geodesic functions on an annulus with $n$  $\mathbb Z_2$-orbifold points into the twisted Yangian. The goal of the present note is to construct a quantum version of this affinnization: given any planar directed network on a disc with its transport matrix $\mathcal M$ of size $(m+n_2)\times (n_1+m)$, the cluster variables of this network parameterize a (finite-dimensional) symplectic leaf in an infinite-dimensional manifold of affine $T(\lambda)$. All matrix entries of $T(\lambda)$ are then polynomials (for an acyclic network) or rational functions (for a network with cycles) of quantum cluster variables. We therefore have an extensive family of possible Poisson submanifolds parameterized by clusters. If we additionally require the upper-right $m\times m$ block of the matrix $\mathcal M$ to be invertible, we can construct the whole quantum loop algebra and the related twisted Yangian. 

All proofs in this paper are rather technical, but otherwise elementary and use only $R$-matrix relations.

\section{A planar network and the quantum algebras of transport matrices}\label{s:intro}
\setcounter{equation}{0}

\subsection{Algebra of matrix elements for the block transport matrix}\label{ss:transport}

We consider a planar network in the disc with $n_1+m$ sources ordered clockwise ($n_1 | m$) and with separated $m+n_2$ sinks ordered counterclockwise ($m | n_2)$. The transport matrix then has the block structure
\be
\label{Mcal}
\mathcal M=\begin{array}{r|c|c|}
 & n_1 & m\\
 \hline
 m & M_{11} & M_{12} \\
 \hline
 n_2 & M_{21} & M_{22} \\
 \hline
 \end{array},
\ee
and we let $M_{\alpha\beta}$, $\alpha,\beta=1,2$, denote its matrix blocks.

{\bf Conventions.} Every $k\times p$ (quantum) matrix $M$ lies in the direct product of $\hbox{Mat}_{k\times p}\otimes Q$, where $Q$ is the set of operators acting in the same quantum space for all matrix elements. We can interpret the matrix $M$ as a morphism from a vector space $V_p\otimes W$ into a vector space $V_k\otimes W$, where $W$ is a (finite- or infinite-dimensional) representation space for  operators from the set $Q$. For cluster-algebra realizations in this paper, $W$ is infinite-dimensional unless $q$ is a root of unity; in fact, $W$ plays no role in the construction below.  In what follows, we assume that the order of product of quantum operators in $Q$ always coincides with the order of the product of  corresponding matrix elements. We also do not indicate explicitly the matrix dimensions of $M_{\alpha\beta}$ and those of $R$-matrices assuming that $R$-matrices are always square matrices whose size is unambiguously determined from a context. All $R$-matrices are assumed to be classical, i.e., they act as the unit operators in $W$. To shorten the writing, we usually omit direct product symbols instead indicating by the index above the symbol of an operator the number of the vector space $V_k$ in which this operator acts nontrivially. For example, for $\bigl[M_{11}\bigr]_{ij}=q_{ij}$
$$
\sheet{1}{M}_{11} = \sheet{1}{M}_{11} \otimes \sheet{2}{\mathbf I}=\sum_k \sum_{i=1}^m \sum_{j=1}^{n_1} \bigl(\sheet{1}{e_{ij}}\otimes \sheet{2} e_{kk}\otimes q_{ij} \bigr),
$$
where we use the standard notation $e_{ij}$ for an elementary matrix whose all elements vanish except the unit element at the intersection of the $i$th row and $j$th column. In what follows, we also use $I$ to denote the unit matrix acting on $V_k$ and $\mathbf I$ to denote the unit operator acting in the direct product $V_k\otimes W$. 

In \cite{ChSh2}, we showed that assuming the Weyl ordering of all path weights in the $(m+n_2)\times (n_1+m)$ matrix $\mathcal M$, it satisfies the quantum $R$-matrix relation (\ref{RMM}), where the $R$-matrix is of size $(m+n_2)^2\times (m+n_2)^2$  in the left-hand side and is of size $(n_1+m)^2\times (n_1+m)^2$ in the right-hand side.
Note that interchanging the spaces $1$ and $2$ results in transposing $R$:
\be
\label{PR}
PR=R^{\text{T}}P,\ \hbox{where}\ P:=\sum_{i,j} \sheet{1}{e}_{ij}\otimes \sheet{2}{e}_{ji}\ \hbox{is the permutation matrix}.
\ee
Besides the standard quantum Yang--Baxter relations, the trigonometric $R$-matrix enjoys special relations:
\be
R^{-1}(q)=R(q^{-1}),
\ee
and
\be
\label{RR-PR}
RR^{\text{T}}=(q{-}q^{-1})RP+I, \ \hbox{or}\ R^{\text{T}}-R^{-1}=R-R^{-\text{T}}=(q{-}q^{-1})P.
\ee

The first statement describes the quantum algebra of $M_{\alpha\beta}$, $\alpha, \beta=1,2$.

\begin{lemma}\label{lm:MM}
Given the block-matrix representation (\ref{Mcal}) and the relation (\ref{RMM}) with the quantum $R$-matrix (\ref{R-matrix1}), the matrix blocks $M_{\alpha\beta}$ satisfy the quantum algebra
\be
\label{Rab}
R\sheet{1}{M}_{\alpha\beta}\sheet{2}{M}_{\alpha\beta}=\sheet{2}{M}_{\alpha\beta}\sheet{1}{M}_{\alpha\beta}R\ \hbox{or}\ R^{-\text{T}}\sheet{1}{M}_{\alpha\beta}\sheet{2}{M}_{\alpha\beta}=\sheet{2}{M}_{\alpha\beta}\sheet{1}{M}_{\alpha\beta}R^{-\text{T}},\quad \alpha\beta=11,12,21,22,
\ee
\begin{align}
&\sheet{2}{M}_{11}\sheet{1}{M}_{12}=R\sheet{1}{M}_{12}\sheet{2}{M}_{11} & \sheet{1}{M}_{12}\sheet{2}{M}_{22}=\sheet{2}{M}_{22}\sheet{1}{M}_{12}R \nonumber \\
\label{M-alg}
&\sheet{1}{M}_{11}\sheet{2}{M}_{21}=\sheet{2}{M}_{21}\sheet{1}{M}_{11}R & \sheet{2}{M}_{21}\sheet{1}{M}_{22}=R\sheet{1}{M}_{22}\sheet{2}{M}_{21} \\
&\sheet{1}{M}_{12}\sheet{2}{M}_{21}=\sheet{2}{M}_{21}\sheet{1}{M}_{12} & \sheet{1}{M}_{11}\sheet{2}{M}_{22}-\sheet{2}{M}_{22}\sheet{1}{M}_{11}=(q{-}q^{-1})\sheet{2}{M}_{21}\sheet{1}{M}_{12}P \nonumber
\end{align}
\end{lemma}

The {\bf proof} is just an evaluation of matrix equalities using the block form of $R$ and $\sheet{1}{\mathcal M}\sheet{2}{\mathcal M}$,
\be\label{R-block}
R=\left[\begin{array}{c|c||c|c} R & 0 & 0 & 0\\ \hline 0 & I &(q{-}q^{-1})P &0\\ \hline\hline 0&0& I &0\\ \hline 0&0&0& R   \end{array}  \right],\quad
\sheet{1}{\mathcal M}\sheet{2}{\mathcal M}=
\left[\begin{array}{c|c||c|c} \sheet{1}{M}_{11}\sheet{2}{M}_{11} & \sheet{1}{M}_{12}\sheet{2}{M}_{11} & \sheet{1}{M}_{11}\sheet{2}{M}_{12} & \sheet{1}{M}_{12}\sheet{2}{M}_{12}\\ \hline \sheet{1}{M}_{21}\sheet{2}{M}_{11} & \sheet{1}{M}_{22}\sheet{2}{M}_{11} &\sheet{1}{M}_{21}\sheet{2}{M}_{12} &\sheet{1}{M}_{22}\sheet{2}{M}_{12}\\ \hline\hline \sheet{1}{M}_{11}\sheet{2}{M}_{21}&\sheet{1}{M}_{12}\sheet{2}{M}_{21}& \sheet{1}{M}_{11}\sheet{2}{M}_{22} &\sheet{1}{M}_{12}\sheet{2}{M}_{22}\\ \hline \sheet{1}{M}_{21}\sheet{2}{M}_{21}&\sheet{1}{M}_{22}\sheet{2}{M}_{21}&\sheet{1}{M}_{21}\sheet{2}{M}_{22}& \sheet{1}{M}_{22}\sheet{2}{M}_{22}   \end{array}  \right]
\ee
and relations (\ref{RR-PR}).

\subsection{Affine Lie--Poisson algebra}

\begin{definition}\label{df:Tk}
We define the {\em level-$k$} quantum transport matrices $T_k$, $k=0,1,2,\dots$ to be
\be
\label{T}
T_0=M_{21},\quad T_1=M_{22}M_{11},\quad T_{k+1}=M_{22}[M_{12}]^k M_{11}, \ k\ge 1.
\ee
\end{definition}

\begin{figure}[h]
\begin{pspicture}(-3.5,-3)(3.5,3){\psset{unit=1.5}
%
\pspolygon[linecolor=red,linewidth=2pt,fillstyle=solid,opacity=0.8,fillcolor=lightgray](-0.3,-1)(-0.3,1)(2.3,1)(2.3,-1)
\psline[doubleline=true,linewidth=1pt, doublesep=1pt, linecolor=black]{->}(2.5,0.75)(2.1,0.75)
\psline[doubleline=true,linewidth=1pt, doublesep=1pt, linecolor=black]{->}(2.5,0.25)(2.1,0.25)
\psline[doubleline=true,linewidth=1pt, doublesep=1pt, linecolor=black]{->}(2.5,-0.25)(2.1,-0.25)
\psline[doubleline=true,linewidth=1pt, doublesep=1pt, linecolor=black]{->}(2.5,-0.75)(2.1,-0.75)
\put(2.7,0){\makebox(0,0)[cl]{\hbox{{$n_1$}}}}
\psline[doubleline=true,linewidth=1pt, doublesep=1pt, linecolor=black]{->}(-0.1,0)(-0.5,0)
\psline[doubleline=true,linewidth=1pt, doublesep=1pt, linecolor=black]{->}(-0.1,0.6)(-0.5,0.6)
\psline[doubleline=true,linewidth=1pt, doublesep=1pt, linecolor=black]{->}(-0.1,-0.6)(-0.5,-0.6)
\put(-0.6,0){\makebox(0,0)[cr]{\hbox{{$n_2$}}}}
\multiput(0,0)(0.4,0){5}{\psline[doubleline=true,linewidth=1pt, doublesep=1pt, linecolor=black]{->}(0.2,0.8)(0.2,1.2)}
%
\multiput(0,0)(0.4,0){5}{\psline[doubleline=true,linewidth=1pt, doublesep=1pt, linecolor=black]{->}(0.2,-1.2)(0.2,-0.8)}
%
\psellipticarc[linecolor=blue,linewidth=2pt,linestyle=dashed](-0.5,1.2)(0.7,0.2){0}{180}
\psellipticarc[linecolor=blue,linewidth=2pt,linestyle=dashed](-0.5,1.2)(1.1,0.3){0}{180}
\psellipticarc[linecolor=blue,linewidth=2pt,linestyle=dashed](-0.5,1.2)(1.5,0.4){0}{180}
\psellipticarc[linecolor=blue,linewidth=2pt,linestyle=dashed](-0.5,1.2)(1.9,0.5){0}{180}
\psellipticarc[linecolor=blue,linewidth=2pt,linestyle=dashed](-0.5,1.2)(2.3,0.6){0}{180}
\psellipticarc[linecolor=blue,linewidth=2pt,linestyle=dashed](-0.5,-1.2)(0.7,0.2){180}{360}
\psellipticarc[linecolor=blue,linewidth=2pt,linestyle=dashed](-0.5,-1.2)(1.1,0.3){180}{360}
\psellipticarc[linecolor=blue,linewidth=2pt,linestyle=dashed](-0.5,-1.2)(1.5,0.4){180}{360}
\psellipticarc[linecolor=blue,linewidth=2pt,linestyle=dashed](-0.5,-1.2)(1.9,0.5){180}{360}
\psellipticarc[linecolor=blue,linewidth=2pt,linestyle=dashed](-0.5,-1.2)(2.3,0.6){180}{360}
\psline[linecolor=blue,linewidth=2pt,linestyle=dashed](-1.2,1.2)(-1.2,-1.2)
\psline[linecolor=blue,linewidth=2pt,linestyle=dashed](-1.6,1.2)(-1.6,-1.2)
\psline[linecolor=blue,linewidth=2pt,linestyle=dashed](-2,1.2)(-2,-1.2)
\psline[linecolor=blue,linewidth=2pt,linestyle=dashed](-2.4,1.2)(-2.4,-1.2)
\psline[linecolor=blue,linewidth=2pt,linestyle=dashed](-2.8,1.2)(-2.8,-1.2)
\put(-2.9,0.2){\makebox(0,0)[br]{\hbox{{$m$}}}}
\psline[linecolor=red,linewidth=2pt,linestyle=dotted](-3,0)(-1,0)

\psline[doubleline=true,linewidth=2pt,linecolor=red,linearc=0.3,doublesep=1.5pt]{->}(0.6,-0.8)(0.6,-0.2)(-0.2,-0.2)
\put(-0.1,-0.7){\makebox(0,0)[bl]{\hbox{{$M_{22}$}}}}
\psline[doubleline=true,linewidth=2pt,linecolor=red,linearc=0.3,doublesep=1.5pt]{<-}(1.4,0.8)(1.4,0.2)(2,0.2)
\put(2.1,0.7){\makebox(0,0)[tr]{\hbox{{$M_{11}$}}}}
\psline[doubleline=true,linewidth=2pt,linecolor=red,linearc=0.3,doublesep=1.5pt]{<-}(-0.2,0.2)(0.6,0.2)(1.4,-0.2)(2,-0.2)
\put(0,0.4){\makebox(0,0)[bl]{\hbox{{$M_{21}$}}}}
\psline[doubleline=true,linewidth=2pt,linecolor=red,linearc=0.3,doublesep=2pt]{->}(1,-0.75)(1,0.75)
\put(1.1,-0.7){\makebox(0,0)[bl]{\hbox{{$M_{12}$}}}}
%
}
\end{pspicture}
	\caption{\small
		A general network $N$ with $n_1+m$ sources and $m+n_2$ sinks with $m$ sources and sinks pairwise identified (dashed lines in the picture). The corresponding boundary cluster variables on the upper and lower sides of the network are then amalgamated into new $m$ ``composite'' cluster variables; paths that contribute to $T_k$ go exactly $k$ times through the cut that separates halves of amalgamated variables in the picture.
	}
	\label{fi:annulus}
\end{figure}
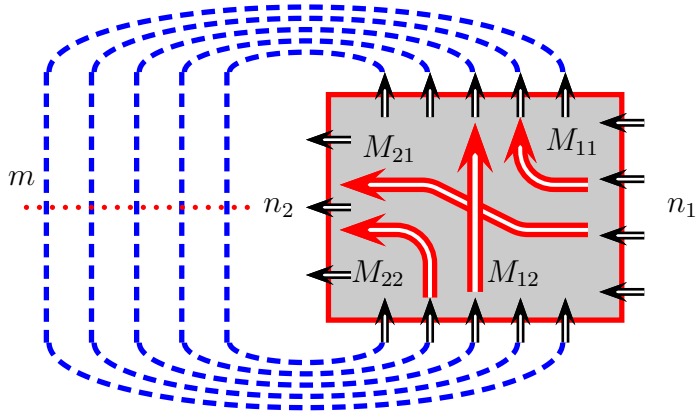

This definition implies that the elements of $T_k$ correspond to paths that cross exactly $k$ times a cut---the dotted line in the Fig.~\ref{fi:annulus}, in which we also indicate paths included into the corresponding matrix blocks. Note also that performing the matrix products in the above definition effectively results in pairwise amalgamations of boundary (frozen) cluster variables on the upper and lower parts of the network in Fig.~\ref{fi:annulus}, i.e, as a result of this operation we obtain a network on an annulus with $m$ threads passing through a cut that separates halves of the amalgamated variables. 

The main theorem of this section follows.
\begin{theorem}\label{th:A}
Quantum transport matrices (\ref{T}) satisfy the quantum algebra
$$
\tcr{R}\sheet{1}T_k\sheet{2}T_p+\tcr{(q{-}q^{-1})P}\sum_{m=1}^p \sheet{1}T_{k+m}\sheet{2}T_{p-m}- \sheet{2}T_p\sheet{1}T_k\tcr{R} - \sum_{m=1}^p \sheet{2}T_{p-m}\sheet{1}T_{k+m}\tcr{(q{-}q^{-1})P}=0\ \hbox{for}\ k\ge p\ge 0.
$$
\end{theorem}

\proof
Assume first that $q,r\ge 0$ and transform $\sheet{1}T_{q+1}\sheet{2}T_{p+1}$ and $\sheet{2}T_{p+1}\sheet{1}T_{q+1}$:
\begin{align*}
\sheet{1}T_{q+1}\sheet{2}T_{p+1}&=\sheet{1}{M}_{22}[\sheet{1}{M}_{12}]^q\tcr{ \sheet{1}{M}_{11}\sheet{2}{M}_{22}}[\sheet{2}{M}_{12}]^p\sheet{2}{M}_{11}\\
&=\sheet{1}{M}_{22}[\sheet{1}{M}_{12}]^q \tcr{ \sheet{2}{M}_{22}}\tcb{\sheet{1}{M}_{11}}[\sheet{2}{M}_{12}]^p\sheet{2}{M}_{11}+(q{-}q^{-1})\sheet{1}{M}_{22}[\sheet{1}{M}_{12}]^q  \sheet{2}{M}_{21}\sheet{1}{M}_{12}\tcr{P} [\sheet{2}{M}_{12}]^p\sheet{2}{M}_{11}\\
&=\sheet{1}{M}_{22}\tcr{\sheet{2}{M}_{22}}[\sheet{1}{M}_{12}R]^q[R^{\text{T}}\sheet{2}{M}_{12}]^p \tcb{\sheet{1}{M}_{11}}\sheet{2}{M}_{11}+
(q{-}q^{-1})\sheet{1}{M}_{22}[\sheet{1}{M}_{12}]^{q+p+1}  \sheet{2}{M}_{21}\sheet{1}{M}_{11}\tcr{P} 
\end{align*}
(Recall that quantum entries of $M_{21}$ commute with those of $M_{12}$.) Correspondingly, 
$$
\sheet{2}T_{p+1}\sheet{1}T_{q+1}=\tcr{\sheet{2}{M}_{22}}\sheet{1}{M}_{22}[\sheet{2}{M}_{12}R^{\text{T}}]^p[R\sheet{1}{M}_{12}]^q \sheet{2}{M}_{11}\tcb{\sheet{1}{M}_{11}}+
(q{-}q^{-1})\tcr{P} \sheet{1}{M}_{22}[\sheet{1}{M}_{12}]^{q+p+1}  \sheet{2}{M}_{21}\sheet{1}{M}_{11}
$$
We now collect in expression in Theorem~\ref{th:A} all terms containing $M_{21}$. For every term $(q{-}q^{-1})P  \sheet{1}T_{k+m}\sheet{2}T_{p-m}-\sheet{2}T_{p-m}\sheet{1}T_{k+m}(q{-}q^{-1})P$ with $m<p$, we have the same combination of operators $\sheet{1}{M}_{22}[\sheet{1}{M}_{12}]^{q+p+1}  \sheet{2}{M}_{21}\sheet{1}{M}_{11}$ sandwiched between two permutation matrices, so this contribution vanishes. It remains only to consider terms with the $R$-matrix insertions and with $m=p$. There, we have, using the formulas above,
\begin{align*}
&(q{-}q^{-1})\tcr{R}\sheet{1}{M}_{22}\sheet{2}{M}_{21}[\sheet{1}{M}_{12}]^{q+p+1}  \sheet{1}{M}_{11}\tcr{P} -(q{-}q^{-1})\tcr{P} \sheet{1}{M}_{22}[\sheet{1}{M}_{12}]^{q+p+1}  \sheet{2}{M}_{21}\sheet{1}{M}_{11}\tcr{R}\\&+(q{-}q^{-1})\tcr{P} \sheet{1}{M}_{22}[\sheet{1}{M}_{12}]^{q+p+1}  \sheet{1}{M}_{11}\sheet{2}{M}_{21}-\sheet{2}{M}_{21}\sheet{1}{M}_{22}[\sheet{1}{M}_{12}]^{q+p+1}  \sheet{1}{M}_{11}(q{-}q^{-1})\tcr{P} \\
=&(q{-}q^{-1})(\tcr{R}\sheet{1}{M}_{22}\sheet{2}{M}_{21}-\sheet{2}{M}_{21}\sheet{1}{M}_{22})[\sheet{1}{M}_{12}]^{q+p+1}  \sheet{1}{M}_{11}\tcr{P}\\
&-(q{-}q^{-1})\tcr{P} \sheet{1}{M}_{22}[\sheet{1}{M}_{12}]^{q+p+1} ( \sheet{2}{M}_{21}\sheet{1}{M}_{11}\tcr{R}- \sheet{1}{M}_{11}\sheet{2}{M}_{21})=0
\end{align*}

We now concentrate on the bulk of terms. Using that $\tcr{R}\sheet{1}{M}_{22}\sheet{2}{M}_{22}=\sheet{2}{M}_{22}\sheet{1}{M}_{22}\tcr{R}$, $\tcr{P}\sheet{1}{M}_{22}\sheet{2}{M}_{22}=\sheet{2}{M}_{22}\sheet{1}{M}_{22}\tcr{P}$, $\sheet{2}{M}_{11}\sheet{1}{M}_{11}\tcr{R}=\tcr{R}\sheet{1}{M}_{11}\sheet{2}{M}_{11}$, and $\sheet{2}{M}_{11}\sheet{1}{M}_{11}\tcr{P}=\tcr{P}\sheet{1}{M}_{11}\sheet{2}{M}_{11}$, we reduce the whole part not containing $M_{21}$ to the expression sandwiched between $\sheet{2}{M}_{22}\sheet{1}{M}_{22}$ and $\sheet{1}{M}_{11}\sheet{2}{M}_{11}$:
\begin{align*}
\sheet{2}{M}_{22}\sheet{1}{M}_{22}\Bigl[ \tcr{R}[\sheet{1}{M}_{12}R]^k[R^{\text{T}}\sheet{2}{M}_{12}]^p+(q{-}q^{-1}) \tcr{P}\sum_{m=1}^p[\sheet{1}{M}_{12}R]^{k+m}[R^{\text{T}}\sheet{2}{M}_{12}]^{p-m}\Bigr.\\
-\Bigl.[\sheet{2}{M}_{12}R^{\text{T}}]^p[R\sheet{1}{M}_{12}]^k  \tcr{R}-(q{-}q^{-1})\sum_{m=1}^p[\sheet{2}{M}_{12}R^{\text{T}}]^{p-m}[R\sheet{1}{M}_{12}]^{k+m} \tcr{P} \Bigr] \sheet{1}{M}_{11}\sheet{2}{M}_{11}
\end{align*}

We now prove that the expression in brackets is zero. We do this in two steps. First, we prove the following lemma,
\begin{lemma}
For $k\ge p\ge 1$,
\begin{align*}
&\tcr{R}[\sheet{1}{M}_{12}R]^k[R^{\text{T}}\sheet{2}{M}_{12}]^p+(q{-}q^{-1}) \tcr{P}\sum_{m=1}^p[\sheet{1}{M}_{12}R]^{k+m}[R^{\text{T}}\sheet{2}{M}_{12}]^{p-m}\\
&-[\sheet{2}{M}_{12}R^{\text{T}}]^p[R\sheet{1}{M}_{12}]^k  \tcr{R}-(q{-}q^{-1})\sum_{m=1}^p[\sheet{2}{M}_{12}R^{\text{T}}]^{p-m}[R\sheet{1}{M}_{12}]^{k+m} \tcr{P} \\
=&\sheet{2}{M}_{12}\sheet{1}{M}_{12}\Bigl[\tcr{R}[\sheet{1}{M}_{12}R]^{k-1}[R^{\text{T}}\sheet{2}{M}_{12}]^{p-1}+(q{-}q^{-1}) \tcr{P}\sum_{m=1}^{p-1}[\sheet{1}{M}_{12}R]^{k+m-1}[R^{\text{T}}\sheet{2}{M}_{12}]^{p-m-1}\Bigr.\\
&-\Bigl.[\sheet{2}{M}_{12}R^{\text{T}}]^{p-1}[R\sheet{1}{M}_{12}]^{k-1}  \tcr{R}-(q{-}q^{-1})\sum_{m=1}^{p-1}[\sheet{2}{M}_{12}R^{\text{T}}]^{p-m-1}[R\sheet{1}{M}_{12}]^{k+m-1} \tcr{P} \Bigr].
\end{align*}
\end{lemma}
We prove this lemma term-by-term: consider first two terms in the sum with the same $1\ge m<p$:
\begin{align*}
&(q{-}q^{-1}) \tcr{P} [\sheet{1}{M}_{12}R]^{k+m}[R^{\text{T}}\sheet{2}{M}_{12}]^{p-m} -(q{-}q^{-1}) [\sheet{2}{M}_{12}R^{\text{T}}]^{p-m}[R\sheet{1}{M}_{12}]^{k+m} \tcr{P}\\
=& (q{-}q^{-1}) \tcr{P} [\sheet{1}{M}_{12}R]^{k+m-1} \sheet{1}{M}_{12} \tcb{[(q{-}q^{-1})RP +I]} \sheet{2}{M}_{12}  [R^{\text{T}}\sheet{2}{M}_{12}]^{p-m-1}\\
&- (q{-}q^{-1}) [\sheet{2}{M}_{12}R^{\text{T}}]^{p-m-1}\sheet{2}{M}_{12}  \tcb{[(q{-}q^{-1})PR +I]}\sheet{1}{M}_{12}[R\sheet{1}{M}_{12}]^{k+m-1} \tcr{P}\\
=& (q{-}q^{-1})^2 [ \tcr{P} [\sheet{1}{M}_{12}R]^{k+p-1} \sheet{1}{M}_{12}\tcb{P} - \tcb{P} [\sheet{1}{M}_{12}R]^{k+p-1} \sheet{1}{M}_{12}\tcr{P}]\\
&+(q{-}q^{-1}) \tcr{P} [\sheet{1}{M}_{12}R]^{k+m-1} \tcb{ \sheet{1}{M}_{12}  \sheet{2}{M}_{12} } [R^{\text{T}}\sheet{2}{M}_{12}]^{p-m-1}-
(q{-}q^{-1}) [\sheet{2}{M}_{12}R^{\text{T}}]^{p-m-1}\tcb{\sheet{2}{M}_{12} \sheet{1}{M}_{12}}[R\sheet{1}{M}_{12}]^{k+m-1} \tcr{P}\\
=&(q{-}q^{-1}) \tcr{P} \tcb{ \sheet{1}{M}_{12}  \sheet{2}{M}_{12} }  [\sheet{1}{M}_{12}R]^{k+m-1} [R^{\text{T}}\sheet{2}{M}_{12}]^{p-m-1}-
(q{-}q^{-1}) \tcb{\sheet{2}{M}_{12} \sheet{1}{M}_{12}} [\sheet{2}{M}_{12}R^{\text{T}}]^{p-m-1}[R\sheet{1}{M}_{12}]^{k+m-1} \tcr{P}\\
=&\tcb{ \sheet{2}{M}_{12}  \sheet{1}{M}_{12} } \Bigl[ (q{-}q^{-1}) \tcr{P}  [\sheet{1}{M}_{12}R]^{k+m-1} [R^{\text{T}}\sheet{2}{M}_{12}]^{p-m-1}- (q{-}q^{-1})
 [\sheet{2}{M}_{12}R^{\text{T}}]^{p-m-1}[R\sheet{1}{M}_{12}]^{k+m-1} \tcr{P}\Bigr].
\end{align*}
And for the remaining four terms, we have
\begin{align*}
&\tcr{R}[\sheet{1}{M}_{12}R]^k[R^{\text{T}}\sheet{2}{M}_{12}]^p+(q{-}q^{-1}) \tcr{P}[\sheet{1}{M}_{12}R]^{k+p}-[\sheet{2}{M}_{12}R^{\text{T}}]^p[R\sheet{1}{M}_{12}]^k  \tcr{R}-(q{-}q^{-1})[R\sheet{1}{M}_{12}]^{k+p} \tcr{P}\\
=&\Bigl[ R[\sheet{1}{M}_{12}R]^{k-1} \sheet{1}{M}_{12}  \tcb{[(q{-}q^{-1})R\tcb{P} +I]} \sheet{2}{M}_{12}  [R^{\text{T}}\sheet{2}{M}_{12}]^{p-1} -(q{-}q^{-1})[R\sheet{1}{M}_{12}]^{k+p} \tcr{P}\Bigr]\\
&-\Bigl[ [\sheet{2}{M}_{12}R^{\text{T}}]^{p-1}\sheet{2}{M}_{12}  \tcb{[(q{-}q^{-1})\tcb{P} R +I]} \sheet{1}{M}_{12} [R\sheet{1}{M}_{12}]^{k-1}  R - (q{-}q^{-1}) \tcr{P}[\sheet{1}{M}_{12}R]^{k+p} \Bigr]\\
=&\Bigl[ (q{-}q^{-1})[R\sheet{1}{M}_{12}]^{k+p} \tcb{P} -(q{-}q^{-1})[R\sheet{1}{M}_{12}]^{k+p}  \tcr{P} + \tcr{R} [\sheet{1}{M}_{12}R]^{k-1}  \tcb{\sheet{1}{M}_{12}  \sheet{2}{M}_{12}}  [R^{\text{T}}\sheet{2}{M}_{12}]^{p-1} \Bigr]\\
&-\Bigl[ [ (q{-}q^{-1}) \tcb{P} [\sheet{1}{M}_{12}R]^{k+p}  - (q{-}q^{-1}) \tcr{P}[\sheet{1}{M}_{12}R]^{k+p} + [\sheet{2}{M}_{12}R^{\text{T}}]^{p-1}  \tcb{\sheet{2}{M}_{12}  \sheet{1}{M}_{12}}  [ R\sheet{1}{M}_{12}]^{k-1}  \tcr{R}  \Bigr]\\
=&\tcr{R}  \tcb{\sheet{1}{M}_{12}  \sheet{2}{M}_{12}}  [\sheet{1}{M}_{12}R]^{k-1}  [R^{\text{T}}\sheet{2}{M}_{12}]^{p-1} - \tcb{\sheet{2}{M}_{12}  \sheet{1}{M}_{12}} [\sheet{2}{M}_{12}R^{\text{T}}]^{p-1}    [ R\sheet{1}{M}_{12}]^{k-1}\tcr{R}\\
=& \sheet{2}{M}_{12}  \sheet{1}{M}_{12} \Bigl[ \tcr{R}  [\sheet{1}{M}_{12}R]^{k-1}  [R^{\text{T}}\sheet{2}{M}_{12}]^{p-1} - [\sheet{2}{M}_{12}R^{\text{T}}]^{p-1}    [ R\sheet{1}{M}_{12}]^{k-1}\tcr{R}\Bigr],
\end{align*}
which completes the proof of the lemma. Using it we can proceed by induction reducing $k\to k-1$ and $p\to p-1$ at every step until we come to $p=1$. Then we have the above four terms for $p=1$ and their combination eventually gives:
\begin{align*}
&\tcr{R}[\sheet{1}{M}_{12}R]^k R^{\text{T}}\sheet{2}{M}_{12}+(q{-}q^{-1}) \tcr{P}[\sheet{1}{M}_{12}R]^{k+1}-\sheet{2}{M}_{12}R^{\text{T}} [R\sheet{1}{M}_{12}]^k  \tcr{R}-(q{-}q^{-1})[R\sheet{1}{M}_{12}]^{k+1} \tcr{P}\\
=&\tcr{R}[\sheet{1}{M}_{12}R]^{k-1} \tcb{\sheet{1}{M}_{12}  \sheet{2}{M}_{12}} - \tcb{\sheet{2}{M}_{12}  \sheet{1}{M}_{12}} R [\sheet{1}{M}_{12}R]^{k-1}
=\bigl[\tcb{R}\tcb{\sheet{1}{M}_{12}  \sheet{2}{M}_{12}}  -   \tcb{\sheet{2}{M}_{12}  \sheet{1}{M}_{12}} [\tcb{R} \bigr] [\sheet{1}{M}_{12}R]^{k-1}=0.
\end{align*}
This completes the proof of the theorem.\quad$\square$

\begin{remark}
The statement of Theorem~\ref{th:A} is very well known in the literature: it is a component-wise form of writing of the celebrated  Lie--Poisson ``$RTT-TTR$'' relation with the spectral parameter. If we introduce
$$
T(\lambda)=\sum_{k=0}^\infty T_k\lambda^{-k},\quad \lambda\in\mathbb C,
$$
then the statement of Theorem~\ref{th:A}  can be written in a compact form as
\be
\label{RmuTT}
R(\lambda,\mu)\sheet{1}T(\lambda)\sheet{2}T(\mu)=\sheet{2}T(\mu)\sheet{1}T(\lambda)R(\lambda,\mu),
\ee
where 
\be
\label{R-affine}
R(\lambda,\mu):= \lambda R^{-\text{T}} - \mu R
\ee
is the trigonometric $R$-matrix with the spectral parameter.
\end{remark}
We therefore have the statement
\begin{theorem}
Quantum cluster variables of any planar network on a disc parameterize a symplectic leaf of the affine Lie--Poisson algebra. 
\end{theorem}

The semiclassical version of this relation for an arbitrary network on an annulus was obtained by Gekhtman, Shapiro, and Vainstein in \cite{GSV3} 

\section{Realizations of quantum loop algebras}\label{s:loop}

We are now about to extend our construction to the case of quantum loop algebras~,\cite{Drinfeld1},\cite{Drinfeld2},\cite{Drinfeld3}.
\begin{definition}\label{def:loop}
The {\em quantum loop algebra} is defined for two sets of matrix-valued elements acting in the direct product of classical and quantum representation spaces $V_k\otimes W$:
\be
T^+(u)=\sum_{k=0}^\infty T^+_k u^{-k}\ \hbox{and} \ T^-(u)=\sum_{k=1}^\infty T^-_k u^{k},\quad u\in\mathbb C,
\ee
subject to the algebra
\begin{align}
\label{R++}
&R(u,v)\sheet{1}T{}^+(u)\sheet{2}T{}^+(v)=\sheet{2}T{}^+(v)\sheet{1}T{}^+(u)R(u,v),\\
\label{R+-}
&R(u,v)\sheet{1}T{}^+(u)\sheet{2}T{}^-(v)=\sheet{2}T{}^-(v)\sheet{1}T{}^+(u)R(u,v),\\
\label{R--}
&R(u,v)\sheet{1}T{}^-(u)\sheet{2}T{}^-(v)=\sheet{2}T{}^-(v)\sheet{1}T{}^-(u)R(u,v).
\end{align}
\end{definition}

An immediate corollary of this definition for the $R$-matrix of form (\ref{R-affine}) is the fourth relation
\be
\label{R-+}
R(u,v)\sheet{1}T{}^-(u)\sheet{2}T{}^+(v)=\sheet{2}T{}^+(v)\sheet{1}T{}^-(u)R(u,v).
\ee

We begin with the remark about invertibility of the quantum matrix $M_{12}$.
\begin{remark}
Relations (\ref{M-alg}) imply that the determinant of the matrix $M_{12}$ in the semiclassical limit has homogeneous commutation relations with all matrix elements of $\mathcal M$; this implies that a proper quantum analogue $\det_q M_{12}$ of this determinant must commute with all entries of $M_{12}$ and $M_{21}$ and must have homogeneous relations $M_{22}\det_q(M_{12})=q \det_q(M_{12})M_{22}$ and $M_{11}\det_q(M_{12})=q^{-1} \det_q(M_{12})M_{11}$. We do not concentrate on details of this inversion operation just assuming that if $\det M_{12}\ne 0$, we can find the quantum inverse operator $M_{12}^{-1}$; the defining relation is that $M_{12}M_{12}^{-1}=\mathbf I$ with $\mathbf I$ being a unit operator in the direct product $V_m\otimes W$ of classical and quantum spaces.
\end{remark}

The main theorem in this section follows.
\begin{theorem}\label{th:loop}
Consider any planar network from Sec.~\ref{ss:transport}. Assume the quantum transport matrix $M_{12}$ to be invertible. Then the elements
\be
\label{TT}
T^+_k:= M_{22} M_{12}^{k-1} M_{11},  \ k\ge 1,\quad T^+_0=M_{21}\ \hbox{and} \ T^-_k:=M_{22} M_{12}^{-k} M_{11} - \delta_{k,1} M_{21},\ k\ge 1,
\ee
satisfy the quantum loop algebra (\ref{R++}--\ref{R--}). In particular, the elements $T^+_0$ and $T^-_1$ constitute a subalgebra
\begin{align}
\label{T0-T0}
&R \sheet{1}T{}^+_0\sheet{2}T{}^+_0= \sheet{2}T{}^+_0\sheet{1}T{}^+_0 R \ \hbox{or} \ R^{-\text{T}} \sheet{1}T{}^+_0\sheet{2}T{}^+_0= \sheet{2}T{}^+_0\sheet{1}T{}^+_0 R^{-\text{T}}, \\
\label{T0-T1}
&R \sheet{1}T{}^-_1\sheet{2}T{}^+_0= \sheet{2}T{}^+_0\sheet{1}T{}^-_1 R, \\
\label{T1-T1}
&R \sheet{1}T{}^-_1\sheet{2}T{}^-_1= \sheet{2}T{}^-_1\sheet{1}T{}^-_1 R \ \hbox{or} \ R^{-\text{T}} \sheet{1}T{}^-_1\sheet{2}T{}^-_1= \sheet{2}T{}^-_1\sheet{1}T{}^-_1 R^{-\text{T}}.
\end{align}
\end{theorem}

the {\bf proof} is a direct (albeit long) calculation. For (\ref{R++}) it is the above Theorem~\ref{th:A}.  Note that in the construction of quantum loop algebra, the commutation relations of zero-level elements  $T^+_0$ and $T^-_1$ between themselves and with all other $T^\pm_k$ play a special role. All these relations must be homogeneous in level indices,
\be
R \sheet{1}T{}^\pm_k\sheet{2}T{}^+_0= \sheet{2}T{}^+_0\sheet{1}T{}^\pm_k R,\qquad R \sheet{1}T{}^-_1\sheet{2}T{}^\pm_k= \sheet{2}T{}^\pm_k\sheet{1}T{}^-_1 R.
\ee
So the first step of the proof is to verify these relations.
It is an easy calculation for $T{}^+_0=M_{21}$ since it commutes with $M_{12}$. It happens however that $M_{22}M_{12}^{-1}M_{11}$ also commutes with $M_{12}$: for this just use the formulas easily derived from (\ref{M-alg}):
\begin{align*}
 &\sheet{2}M_{22} R^{-1}  \sheet{1}M{}_{12}^{-1}= \sheet{1}M{}_{12}^{-1} \sheet{2}M{}_{22},\\
 &\sheet{1}M{}_{12}^{-1} R^{-1}  \sheet{2}M{}_{11}= \sheet{2}M{}_{11} \sheet{1}M{}_{12}^{-1},\\
  &\sheet{1}M{}_{12}^{-1}  \sheet{2}M{}_{12}^{-1} R= R \sheet{2}M{}_{12}^{-1} \sheet{1}M{}_{12}^{-1}.
\end{align*}
Then we have
$$
\sheet{1}M{}_{12}^{-1} \sheet{2}M{}_{22} \sheet{2}M{}_{12}^{-1}   \sheet{2}M{}_{11}= \sheet{2}M_{22} R^{-1}  \sheet{1}M{}_{12}^{-1} \sheet{2}M{}_{12}^{-1}   \sheet{2}M{}_{11} =\sheet{2}M_{22} \sheet{2}M{}_{12}^{-1} \sheet{1}M{}_{12}^{-1}  R^{-1}   \sheet{2}M{}_{11}=\sheet{2}M_{22} \sheet{2}M{}_{12}^{-1}   \sheet{2}M{}_{11} \sheet{1}M{}_{12}^{-1}.
$$
Analogously, we obtain
\be
R \sheet{1}M_{22} \sheet{1}M{}_{12}^{-1}   \sheet{1}M{}_{11}  \sheet{2}M_{22} =  \sheet{2}M_{22} \sheet{1}M_{22} \sheet{1}M{}_{12}^{-1}   \sheet{1}M{}_{11} + (q{-}q^{-1}) R \sheet{1}M_{22} \sheet{2}M_{21}
\ee
and 
\be
\label{XX}
\sheet{1}M_{22} \sheet{1}M{}_{12}^{-1}   \sheet{1}M{}_{11}  \sheet{2}M_{11} R^{-1} =  \sheet{2}M_{11} \sheet{1}M_{22} \sheet{1}M{}_{12}^{-1}   \sheet{1}M{}_{11}  - (q{-}q^{-1})  \sheet{1}M_{21} \sheet{2}M_{11}P,
\ee
so that, we have 
\begin{align*}
&R \sheet{1}M_{22} \sheet{1}M{}_{12}^{-1}   \sheet{1}M{}_{11}  \sheet{2}M_{22}  \sheet{2}M{}^k_{12} \sheet{2}M_{11} 
=\sheet{2}M_{22}  \sheet{1}M_{22} \sheet{1}M{}_{12}^{-1}   \sheet{1}M{}_{11}  \sheet{2}M{}^k_{12} \sheet{2}M_{11} + (q{-}q^{-1}) R \sheet{1}M_{22} \sheet{2}M_{21} P  \sheet{2}M{}^k_{12} \sheet{2}M_{11} \\
=& \sheet{2}M_{22}  \sheet{2}M{}^k_{12} \sheet{2}M_{11}  \sheet{1}M_{22} \sheet{1}M{}_{12}^{-1}   \sheet{1}M{}_{11} R - (q{-}q^{-1})  \sheet{2}M_{22}  \sheet{2}M{}^k_{12} \sheet{1}M_{21} \sheet{2}M_{11} P R   + (q{-}q^{-1}) R P \sheet{2}M_{22} \sheet{1}M_{21} \sheet{2}M{}^k_{12} \sheet{2}M_{11}.
\end{align*}
Using now that $(q{-}q^{-1}) R P = R R^{\text{T}}-I$ and $(q{-}q^{-1}) P R  = R^{\text{T}} R -I$, that the terms proportional to $I$ are mutually cancelled, and that  
$R R^{\text{T}} \sheet{2}M_{22} \sheet{1}M_{21} = R  \sheet{1}M_{21}  \sheet{2}M_{22}$ and $\sheet{1}M_{21} \sheet{2}M_{11} R^{\text{T}}  R=\sheet{2}M_{11}\sheet{1}M_{21} R$, we finally obtain the desired equality
$$
R \bigl(\sheet{1}M_{22} \sheet{1}M{}_{12}^{-1}  \sheet{1}M{}_{11} - \sheet{1}M_{21} \bigr) \sheet{2}M_{22}  \sheet{2}M{}^k_{12} \sheet{2}M_{11} =
\sheet{2}M_{22}  \sheet{2}M{}^k_{12} \sheet{2}M_{11} \bigl(\sheet{1}M_{22} \sheet{1}M{}_{12}^{-1}  \sheet{1}M{}_{11} - \sheet{1}M_{21} \bigr) R.
$$

Although the proof of (\ref{R--}) differs in small details, it is generally close to that of (\ref{R++}) being also rather lengthy; we therefore omit it here. At the same time, as an example of this technique, we evaluate commutation relations between matrix elements of $T^-_2$ in the appendix.

We now present a complete proof of relation (\ref{R+-}).  This relation is equivalent to the relation on $T$'s:
\be \label{rel-kp}
R^{-\text{T}} \sheet{1}T{}^+_{k+1}  \sheet{2}T{}^-_{p} - R  \sheet{1}T{}^+_{k}  \sheet{2}T{}^-_{p-1}-\bigl(   \sheet{2}T{}^-_{p} \sheet{1}T{}^+_{k+1} R^{-\text{T}} -  \sheet{2}T{}^-_{p-1}  \sheet{1}T{}^+_{k}  R\bigr)=0.
\ee
Commuting an inner pair of matrix operators $M_{11}$ and $M_{22}$ in $T$-operators, observing that the resulting terms containing $M_{21}$ are cancelled, and pushing in the obtained operator products the operators $M_{11}$ to the right and $M_{22}$ to the left from their central positions through products of operators $ \sheet{1}M_{12}$ and $ \sheet{2}M{}^{-1}_{12}$, on the first step we reduce the combination of operators (\ref{rel-kp}) to the expression
\begin{align}
&R^{-\text{T}} \bigl[ \sheet{1}M_{12} R \bigr]^k  \bigl[ \sheet{2}M{}_{12}^{-1} R^{-\text{T}} \bigr]^p - R \bigl[ \sheet{1}M_{12} R \bigr]^{k-1}  \bigl[ \sheet{2}M{}_{12}^{-1} R^{-\text{T}} \bigr]^{p-1}\nonumber\\
&-  \bigl[ R^{-\text{T}}  \sheet{2}M{}_{12}^{-1} \bigr]^p \bigl[ R \sheet{1}M_{12} \bigr]^k R^{-\text{T}} +   \bigl[ R^{-\text{T}}  \sheet{2}M{}_{12}^{-1} \bigr]^{p-1} \bigl[ R \sheet{1}M_{12} \bigr]^{k-1} R
\label{rec-1}
\end{align}
sandwiched between $ \sheet{2}M_{22} \sheet{1}M_{22}$ and $ \sheet{1}M_{11} \sheet{2}M_{11}$. We have therefore to prove that (\ref{rec-1}) vanishes for all $k$ and $p$. We prove this statement by induction. First, it is an easy calculation to see that it holds for $k=1$ and any $p$ and for $p=1$ and any $k$.

Since
$$
\sheet{1}M_{12} R  \sheet{2}M{}_{12}^{-1}= \sheet{2}M{}_{12}^{-1} R \sheet{1}M_{12}\quad \hbox{and} \quad
\sheet{1}M_{12} R^{-\text{T}}   \sheet{2}M{}_{12}^{-1}= \sheet{2}M{}_{12}^{-1} R^{-\text{T}}  \sheet{1}M_{12},
$$ 
in the above expressions we can freely interchange any pair of matrices $\sheet{1}M_{12}$ and $\sheet{2}M{}_{12}^{-1}$ in the string of operators $\cdots M_\bullet R^\bullet M_\bullet R^\bullet M_\bullet \cdots$ provided $M_\bullet$ are either  $\sheet{1}M_{12}$ or $ \sheet{2}M{}_{12}^{-1}$ and they are interlaced with $R^\bullet$ that are either $R$ or $R^{-\text{T}}$. Exploiting this freedom alone we can bring (\ref{rec-1}) to a more convenient form
\begin{align}
&R^{-\text{T}}  \sheet{2}M{}_{12}^{-1} \Bigl[ \tcb{R} \bigl[ \sheet{1}M_{12} R \bigr]^{k-1}  \bigl[ \sheet{2}M{}_{12}^{-1} R^{-\text{T}} \bigr]^{p-1} 
-  \bigl[ R^{-\text{T}}  \sheet{2}M{}_{12}^{-1} \bigr]^{p-1} \bigl[ R \sheet{1}M_{12} \bigr]^{k-1} \tcb{R} \Bigr]   \sheet{1}M_{12} R^{-\text{T}} \nonumber\\
&- \Bigl[ R \bigl[ \sheet{1}M_{12} R \bigr]^{k-1}  \bigl[ \sheet{2}M{}_{12}^{-1} R^{-\text{T}} \bigr]^{p-1} 
-  \bigl[ R^{-\text{T}}  \sheet{2}M{}_{12}^{-1} \bigr]^{p-1} \bigl[ R \sheet{1}M_{12} \bigr]^{k-1} R \Bigr],
\label{rec-2}
\end{align}
where expressions in the brackets in the first and second lines are identical.  For the first line, we transform the two indicated $R$-matrices using that $R=R^{-\text{T}} +(q{-}q^{-1})P$, interchange the rightmost $\sheet{2}M{}_{12}^{-1} $ in the second term with the second (or last, doesn't matter) $ \sheet{1}M_{12}$ on the right and use that 
$$
R^{-\text{T}}  \sheet{2}M{}_{12}^{-1} (q{-}q^{-1})P \sheet{1}M_{12} R = R  \sheet{2}M{}_{12}^{-1}  (q{-}q^{-1})P \sheet{1}M_{12} R^{-\text{T}} =(q{-}q^{-1})P=R-R^{-\text{T}}.
$$
The first line of (\ref{rec-2}) then becomes
\begin{align*}
&R^{-\text{T}}  \sheet{2}M{}_{12}^{-1} R^{-\text{T}}\tcr{ \sheet{1}M_{12}} R   \bigl[ \sheet{1}M_{12} R \bigr]^{k-2}  \bigl[ \sheet{2}M{}_{12}^{-1} R^{-\text{T}} \bigr]^{p-2} \tcr{\sheet{2}M{}_{12}^{-1} }R^{-\text{T}}  \sheet{1}M_{12} R^{-\text{T}} \\
&+(\tcr{R}-R^{-\text{T}}) \bigl[ \sheet{1}M_{12} R \bigr]^{k-2} \tcr{\sheet{2}M{}_{12}^{-1} }  R^{-\text{T}} \bigl[ \sheet{2}M{}_{12}^{-1} R^{-\text{T}} \bigr]^{p-2} \tcr{\sheet{1}M_{12} }R^{-\text{T}}\\
&-R^{-\text{T}}  \sheet{2}M{}_{12}^{-1} R^{-\text{T}}  \sheet{2}M{}_{12}^{-1}  \bigl[ R^{-\text{T}} \sheet{2}M{}_{12}^{-1}  \bigr]^{p-2}  \bigl[ R \sheet{1}M_{12} \bigr]^{k-2} R   \sheet{1}M_{12} R^{-\text{T}}  \sheet{1}M_{12} R^{-\text{T}} \\
&- \bigl[ R^{-\text{T}} \sheet{2}M{}_{12}^{-1}  \bigr]^{p-1} R^{-\text{T}} \sheet{1}M_{12} \bigl[ R \sheet{1}M_{12} \bigr]^{k-2} (R-R^{-\text{T}}) 
\end{align*}
Combining the first and the third terms and expanding the second and the fourth, we obtain
\begin{align*}
&R^{-\text{T}}  \sheet{2}M{}_{12}^{-1} R^{-\text{T}} \sheet{2}M{}_{12}^{-1} \Bigl[  R   \bigl[ \sheet{1}M_{12} R \bigr]^{k-2}  \bigl[ \sheet{2}M{}_{12}^{-1} R^{-\text{T}} \bigr]^{p-2}-\bigl[ R^{-\text{T}} \sheet{2}M{}_{12}^{-1}  \bigr]^{p-2}  \bigl[ R \sheet{1}M_{12} \bigr]^{k-2} R  \Bigr] \sheet{1}M_{12} R^{-\text{T}}  \sheet{1}M_{12} R^{-\text{T}} \\
&+R \bigl[ \sheet{1}M_{12} R \bigr]^{k-2}  \sheet{1}M_{12}\tcb{R^{-\text{T}} } \bigl[ \sheet{2}M{}_{12}^{-1} R^{-\text{T}} \bigr]^{p-1}\\
&-R^{-\text{T}}\tcr{\sheet{2}M{}_{12}^{-1} } R  \bigl[ \sheet{1}M_{12} R \bigr]^{k-3} \tcr{ \sheet{1}M_{12} }\tcb{R^{-\text{T}} }\bigl[ \sheet{2}M{}_{12}^{-1} R^{-\text{T}} \bigr]^{p-2} \sheet{1}M_{12}R^{-\text{T}}\\
&- \bigl[ R^{-\text{T}} \sheet{2}M{}_{12}^{-1}  \bigr]^{p-1}\tcb{ R^{-\text{T}} }\sheet{1}M_{12} \bigl[ R \sheet{1}M_{12} \bigr]^{k-2} R\\
&+ R^{-\text{T}} \sheet{2}M{}_{12}^{-1}  \bigl[ R^{-\text{T}} \sheet{2}M{}_{12}^{-1}  \bigr]^{p-2}\tcb{ R^{-\text{T}} }\bigl[ \sheet{1}M_{12} R \bigr]^{k-2} \sheet{1}M_{12}  R^{-\text{T}}\\
=&R^{-\text{T}}  \sheet{2}M{}_{12}^{-1} R^{-\text{T}} \sheet{2}M{}_{12}^{-1} \Bigl[  R   \bigl[ \sheet{1}M_{12} R \bigr]^{k-2}  \bigl[ \sheet{2}M{}_{12}^{-1} R^{-\text{T}} \bigr]^{p-2}-\bigl[ R^{-\text{T}} \sheet{2}M{}_{12}^{-1}  \bigr]^{p-2}  \bigl[ R \sheet{1}M_{12} \bigr]^{k-2} R  \Bigr] \sheet{1}M_{12} R^{-\text{T}}  \sheet{1}M_{12} R^{-\text{T}} \\
&-R^{-\text{T}} \sheet{2}M{}_{12}^{-1} \Bigl[  R   \bigl[ \sheet{1}M_{12} R \bigr]^{k-2}  \bigl[ \sheet{2}M{}_{12}^{-1} R^{-\text{T}} \bigr]^{p-2}-\bigl[ R^{-\text{T}} \sheet{2}M{}_{12}^{-1}  \bigr]^{p-2}  \bigl[ R \sheet{1}M_{12} \bigr]^{k-2} R  \Bigr] \sheet{1}M_{12} R^{-\text{T}} \\
&+\Bigl[ R \bigl[ \sheet{1}M_{12} R \bigr]^{k-1}  \bigl[ \sheet{2}M{}_{12}^{-1} R^{-\text{T}} \bigr]^{p-1} 
-  \bigl[ R^{-\text{T}}  \sheet{2}M{}_{12}^{-1} \bigr]^{p-1} \bigl[ R \sheet{1}M_{12} \bigr]^{k-1} R \Bigr]\\
&-\bigl[ R \sheet{1}M_{12}  \bigr]^{k-1}  \tcb{(q{-}q^{-1})P} \bigl[ \sheet{2}M{}_{12}^{-1} R^{-\text{T}} \bigr]^{p-1}
+R^{-\text{T}}\sheet{2}M{}_{12}^{-1}   \bigl[ R \sheet{1}M_{12}  \bigr]^{k-2}  \tcb{(q{-}q^{-1})P} \bigl[ \sheet{2}M{}_{12}^{-1} R^{-\text{T}} \bigr]^{p-2} \sheet{1}M_{12}R^{-\text{T}}\\
&+\bigl[ R^{-\text{T}} \sheet{2}M{}_{12}^{-1}  \bigr]^{p-1}\tcb{ (q{-}q^{-1})P } \bigl[ \sheet{1}M_{12} R \bigr]^{k-1}
- R^{-\text{T}} \sheet{2}M{}_{12}^{-1}  \bigl[ R^{-\text{T}} \sheet{2}M{}_{12}^{-1}  \bigr]^{p-2}\tcb{ (q{-}q^{-1})P}\bigl[ \sheet{1}M_{12} R \bigr]^{k-2} \sheet{1}M_{12}  R^{-\text{T}}
\end{align*}
In the resulting expression, the first two lines reproduce relation (\ref{rec-2}) with $k\to k-1$ and $p\to p-1$ sandwiched between $R^{-\text{T}}  \sheet{2}M{}_{12}^{-1}$ and $\sheet{1}M_{12} R^{-\text{T}} $, which provides the recursion implication, the third line is the second line of (\ref{rec-2}), and in order to complete the proof it remains only to show that the expression in the last two lines vanishes identically. Let $k\ge p$. Then, for instance, 
$$
\bigl[ R^{-\text{T}} \sheet{2}M{}_{12}^{-1}  \bigr]^{p-1} (q{-}q^{-1})P  \bigl[ \sheet{1}M_{12} R \bigr]^{k-1}= (q{-}q^{-1})P  \bigl[ \sheet{1}M_{12} R \bigr]^{k-p},
$$
so the mentioned last two lines become
\begin{align*}
&-\bigl[ R \sheet{1}M_{12}  \bigr]^{k-p}  (q{-}q^{-1})P 
+R^{-\text{T}}\sheet{2}M{}_{12}^{-1}   \bigl[ R \sheet{1}M_{12}  \bigr]^{k-p}  (q{-}q^{-1})P \sheet{1}M_{12}R^{-\text{T}}\\
&+ (q{-}q^{-1})P  \bigl[ \sheet{1}M_{12} R \bigr]^{k-p}
- R^{-\text{T}} \sheet{2}M{}_{12}^{-1} (q{-}q^{-1})P\bigl[ \sheet{1}M_{12} R \bigr]^{k-p} \sheet{1}M_{12}  R^{-\text{T}}\\
=&-\bigl[ R \sheet{1}M_{12}  \bigr]^{k-p} (R-R^{-\text{T}}) 
+R^{-\text{T}} \bigl[  \sheet{1}M_{12} R  \bigr]^{k-p} \sheet{2}M{}_{12}^{-1}    (q{-}q^{-1})P \sheet{1}M_{12}R^{-\text{T}}\\
&+ (R-R^{-\text{T}})  \bigl[ \sheet{1}M_{12} R \bigr]^{k-p}
- R^{-\text{T}} \sheet{2}M{}_{12}^{-1} (q{-}q^{-1})P\bigl[ \sheet{1}M_{12} R \bigr]^{k-p} \sheet{1}M_{12}  R^{-\text{T}}\\
=&\bigl[ R \sheet{1}M_{12}  \bigr]^{k-p} R^{-\text{T}}
+R^{-\text{T}} \bigl[  \sheet{1}M_{12} R  \bigr]^{k-p-1}  \sheet{1}M_{12}     (q{-}q^{-1})P\\
&-R^{-\text{T}}  \bigl[ \sheet{1}M_{12} R \bigr]^{k-p}
-  (q{-}q^{-1})P  \sheet{1}M_{12}  \bigl[ R \sheet{1}M_{12} \bigr]^{k-p-1} R^{-\text{T}}\\
=&\bigl[ R \sheet{1}M_{12}  \bigr]^{k-p} R^{-\text{T}}
+R^{-\text{T}} \bigl[  \sheet{1}M_{12} R  \bigr]^{k-p-1}  \sheet{1}M_{12}  (R-R^{-\text{T}})\\
&-R^{-\text{T}}  \bigl[ \sheet{1}M_{12} R \bigr]^{k-p}
-  (R-R^{-\text{T}})  \sheet{1}M_{12}  \bigl[ R \sheet{1}M_{12} \bigr]^{k-p-1} R^{-\text{T}}=0.
\end{align*}
By induction in $k$ and $p$ we therefore conclude that expression (\ref{rec-2}) is identically zero for all $k$ and $p$. The theorem is proved.$\quad\square$

\begin{remark}\label{rm:groupoid}
Note that the combination $T^-_0=M_{22} M_{12}^{-1} M_{11} - M_{21}$ vanishes for a number of interesting planar acyclic networks (see Example~\ref{ex:network} below). 
This is in line with the ``groupoid condition'' \cite{CMR},\cite{ChSh2} appearing in the cases where we can identify transport matrices with monodromy matrices of an $SL_n$ character variety: if this is the case, a natural condition is that monodromy elements corresponding to two homotopically equivalent paths must coincide. To define this equivalence, we must be able to invert directions in a directed network. To satisfy the homotopy condition, transport matrices $T_{\rightarrow}$ and $T_{\leftarrow}$ corresponding to two oppositely directed paths joining boundary components must be mutually inverse: $T_{\rightarrow}\cdot T_{\leftarrow}={\mathbb I}$. Then, say, transport matrices $M_{22} M_{12}^{-1} M_{11}$ and $M_{21}$ correspond to homotopy equivalent paths, and satisfaction of the groupoid condition requires these two matrices to be equal. Note that the groupoid condition is satisfied only for special networks (notably, those of geometric type) and requires a ``fine tuning'' by special $q$-factor matrices (see \cite{ChSh2}). 
\end{remark}

If the groupoid condition is satisfied, we should adjust the statement of Theorem~\ref{th:loop}:
\begin{theorem}\label{th:loop1}
Consider any planar acyclic network from Sec.~\ref{ss:transport}. Assume the quantum transport matrix $M_{12}$ to be invertible and the quantum transport matrices to satisfy the ``groupoid condition'' $M_{21}=M_{22}M_{12}^{-1} M_{11} $. Then the elements
\be
\label{TT1}
T^+_k:= M_{22} M_{12}^{k-1} M_{11},  \ k\ge 0,\ \hbox{and} \ T^-_k:=M_{22} M_{12}^{-k-1} M_{11},\ k\ge 1,
\ee
satisfy the quantum loop algebra (\ref{R++}--\ref{R--}), and the newly defined elements $T^+_0$ and $T^-_1$ constitute a subalgebra (\ref{T0-T0})--(\ref{T1-T1}).
\end{theorem}

\begin{example}\label{ex:network}
Let us consider a planar network with $n_1+(m_2+m_1)$ sources and $(m_2+m_1)+n_2$ sinks that is amalgamated out of three planar subnetworks: the first with $n_1+m_2$ sources and $m_2$ sinks, the second with $m_2+m_1$ sources and $m_2+k$ sinks ($k\ge m_1$ for the matrix $T_1^3 T_3^2$ to be of the full rank) and the third with $k$ sources and $m_1+n_2$ sinks as shown in the picture. Let $T_i^j$ be the corresponding transport matrices (all directed from sources to sinks).
$$
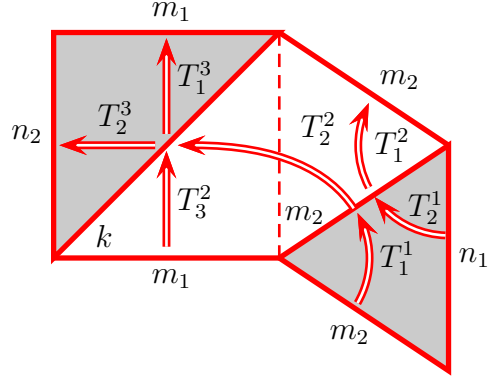
\begin{pspicture}(-3.5,-1.5)(3.5,3.5){\psset{unit=1.5}
%
\pspolygon[linecolor=red,linewidth=2pt,fillstyle=solid,opacity=0.8,fillcolor=lightgray](-2,0)(0,2)(-2,2)
\pspolygon[linecolor=red,linewidth=2pt,fillstyle=solid,opacity=0.8,fillcolor=lightgray](0,0)(1.5,1)(1.5,-1)
\psline[linecolor=red,linewidth=2pt]{-}(-2,0)(0,0)
\psline[linecolor=red,linewidth=2pt]{-}(0,2)(1.5,1)
\psline[linecolor=red,linewidth=1pt,linestyle=dashed]{-}(0,2)(0,0)
\put(0,0){\psline[doubleline=true,linewidth=1pt, doublesep=1pt, linecolor=red]{->}(-1,0.1)(-1,0.95)}
\put(0,0){\psline[doubleline=true,linewidth=1pt, doublesep=1pt, linecolor=red]{->}(-1,1.1)(-1,1.95)}
\put(0,0){\psline[doubleline=true,linewidth=1pt, doublesep=1pt, linecolor=red]{->}(-1.1,1)(-1.95,1)}
\psarc[doubleline=true,linewidth=1pt, doublesep=1pt, linecolor=red]{->}(0,0){0.8}{-30}{30}
\psarc[doubleline=true,linewidth=1pt, doublesep=1pt, linecolor=red]{<-}(1.5,1){0.8}{215}{268}
\psarc[doubleline=true,linewidth=1pt, doublesep=1pt, linecolor=red]{<-}(1.5,1){0.8}{152}{208}
\psbezier[doubleline=true,linewidth=1pt, doublesep=1pt, linecolor=red]{<-}(-0.9,1)(0,1)(0.44,0.77)(0.66,0.44)
\put(-2.1,1.1){\makebox(0,0)[cr]{\hbox{{ $n_2$}}}}
\put(1.5,0){\makebox(0,0)[cl]{\hbox{{ $n_1$}}}}
\put(-1,2.1){\makebox(0,0)[bc]{\hbox{{ $m_1$}}}}
\put(-1,-0.1){\makebox(0,0)[tc]{\hbox{{ $m_1$}}}}
\put(-1.6,0.1){\makebox(0,0)[bc]{\hbox{{ $k$}}}}
\put(0.8,-0.6){\makebox(0,0)[tr]{\hbox{{ $m_2$}}}}
\put(0.8,1.5){\makebox(0,0)[bl]{\hbox{{ $m_2$}}}}
\put(0.4,0.3){\makebox(0,0)[br]{\hbox{{ $m_2$}}}}
\put(1.05,0.3){\makebox(0,0)[bl]{\hbox{{ $T^1_2$}}}}
\put(0.8,0){\makebox(0,0)[cl]{\hbox{{ $T^1_1$}}}}
\put(0.75,1){\makebox(0,0)[cl]{\hbox{{ $T^2_1$}}}}
\put(0.1,1){\makebox(0,0)[bl]{\hbox{{ $T^2_2$}}}}
\put(-1,0.5){\makebox(0,0)[cl]{\hbox{{ $T^2_3$}}}}
\put(-1,1.6){\makebox(0,0)[cl]{\hbox{{ $T^3_1$}}}}
\put(-1.5,1.1){\makebox(0,0)[bc]{\hbox{{ $T^3_2$}}}}
}
\end{pspicture}
$$
We also assume that no transport occurs from the side with $m_1$ sources to the side with $m_2$ sinks in the middle subnetwork.  The total transport matrix $\mathcal M$ then has the form
\be
\label{M-ex}
\mathcal M=\begin{array}{r|c|c|c|}
 & n_1 & m_2& m_1\\
 \hline
 \phantom{\Bigm|} m_2 & T_1^2 T_2^1 & T^2_1 T^1_1 & 0 \\
 \hline
 \phantom{\Bigm|} m_1 & T_1^3 T_2^2 T_2^1 & T_1^3 T_2^2 T_1^1 & T_1^3 T_3^2 \\
 \hline
  \phantom{\Bigm|} n_2 & T_2^3 T_2^2 T_2^1 & T_2^3 T^2_2 T^1_1 & T_2^3 T_3^2 \\
 \hline
 \end{array},
\ee
and in order to ensure the invertibility of $M_{12}$ we have to require that $T^2_1 T^1_1$ and $T_1^3 T_3^2$ both be invertible. The matrix $M_{12}^{-1}$ then reads
$$
M^{-1}_{12}=\begin{array}{r|c|c|}
 & m_2& m_1\\
 \hline
 \phantom{\Bigm|} m_2 &  [T^2_1 T^1_1]^{-1} & 0 \\
 \hline
 \phantom{\Bigm|} m_1 & -  [T_1^3 T_3^2]^{-1} T_1^3 T_2^2 T_1^1  [T^2_1 T^1_1]^{-1} & [T_1^3 T_3^2]^{-1} \\
 \hline
 \end{array}.
$$
We now impose an {\em additional condition} that $T_1^1$ and $T^2_1$ be invertible themselves. Note that this condition holds for virtually all geometric systems relating the networks in question to character varieties of $SL_n$ systems on polygons (discs with $s\ge 3$ marked points on the boundary) in the Fock--Goncharov description of higher Teichm\"uller spaces \cite{FG1}. The combination $M_{22}M_{12}^{-1}M_{11}$ then becomes
\begin{align*}
M_{22}M_{12}^{-1}M_{11}=& T_2^3 T^2_2 (T^1_1  [T^2_1 T^1_1]^{-1} T_1^2 ) T_2^1 - T_2^3 T_3^2 [T_1^3 T_3^2]^{-1} T_1^3 T_2^2 ( T_1^1  [T^2_1 T^1_1]^{-1} T_1^2 ) T_2^1 \\
&\quad + T_2^3 T_3^2 [T_1^3 T_3^2]^{-1} T_1^3 T_2^2 T_2^1 \\
=&T_2^3 T^2_2 T_2^1  - T_2^3 T_3^2 [T_1^3 T_3^2]^{-1} T_1^3 T_2^2  T_2^1 + T_2^3 T_3^2 [T_1^3 T_3^2]^{-1} T_1^3 T_2^2 T_2^1\\
=&T_2^3 T^2_2 T_2^1 = M_{21},
\end{align*}
and the groupoid condition is therefore satisfied for any such network.
\end{example}

\begin{example}\label{ex:network2}
Let us consider a planar network corresponding to a special polygon composed out of $2r$ elementary triangular subnetworks in each of which we assume that all transport matrices are invertible so that all triangles have exactly $n$ sources/sinks on every side; this happens, e.g., for systems corresponding to $SL_n$ higher Teichm\"uller spaces on a disc with $2r+2$ marked points on the boundary; every triangular subnetwork with $n$ sources and $2n$ sinks then has the form depicted in Fig.~\ref{fi:triangle} and every triangular network with $2n$ sources and $n$ sinks can be obtained from the network in  Fig.~\ref{fi:triangle} by inverting all vertical double arrows and interchanging colors of all black and gray vertices
(we should then also invert directions of all edges in the quiver).
$$
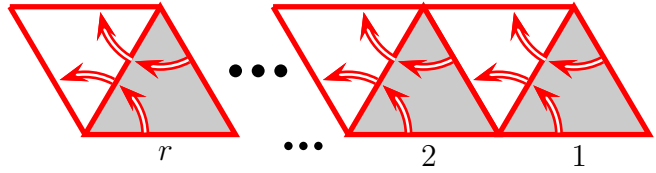
\begin{pspicture}(-3.5,-1.5)(3.5,1.5){\psset{unit=1}
\newcommand{\PATGEN}{%
{\psset{unit=1}
\pspolygon[linecolor=red,linewidth=2pt,fillstyle=solid,opacity=0.8,fillcolor=lightgray](-1,-0.5)(0,1.2)(1,-0.5)
\psline[linecolor=red,linewidth=2pt]{-}(-1,-0.5)(-2,1.2)
\psline[linecolor=red,linewidth=2pt]{-}(-2,1.2)(0,1.2)
\psarc[doubleline=true,linewidth=1pt, doublesep=1pt, linecolor=red]{->}(-1,-0.5){0.8}{2}{55}
\psarc[doubleline=true,linewidth=1pt, doublesep=1pt, linecolor=red]{<-}(0,1.2){0.8}{245}{298}
\psarc[doubleline=true,linewidth=1pt, doublesep=1pt, linecolor=red]{<-}(0,1.2){0.8}{185}{242}
\psarc[doubleline=true,linewidth=1pt, doublesep=1pt, linecolor=red]{->}(-1,-0.5){0.8}{62}{115}
}}
\rput(3,0){\PATGEN}
\rput(1,0){\PATGEN}
\multiput(-0.9,0.35)(-0.3,0){3}{\pscircle[linecolor=white,fillstyle=solid,fillcolor=black](0,0){0.1}}
\multiput(-0.4,-0.65)(-0.2,0){3}{\pscircle[linecolor=white,fillstyle=solid,fillcolor=black](0,0){0.07}}
\rput(-2.5,0){\PATGEN}
\put(3,-0.65){\makebox(0,0)[tc]{\hbox{{ $1$}}}}
\put(1,-0.65){\makebox(0,0)[tc]{\hbox{{ $2$}}}}
\put(-2.5,-0.65){\makebox(0,0)[tc]{\hbox{{ $r$}}}}
}
\end{pspicture}
$$
For this network, we have 
\be
M_{22}\bigl[M_{12}^{-1}\bigr]^p M_{11}=\bigl\{ M_{21}\ \hbox{for} \ p=1,\ \ 0 \ \hbox{for} \ 2\le p\le r,\ \ \hbox{nonzero otherwise}\bigr\}.
\ee
The complete proof can be done by induction in $r$; we however consider only a toy situation in which we replace all $(n\times n)$-blocks of the total $n(r+1)\times n(r+1)$ matrix $\mathcal M$ by units, then $\mathcal M\to \widehat M$, where $\widehat M$ is the $(r+1)\times (r+1)$ matrix with entries  $[\widehat M]_{i,j}$ equal $1$ for $i-j+1\ge 0$ and zero otherwise. The transport matrix $\widehat M_{12}$ is the $r\times r$ unipotent triangular matrix with all entries on the diagonal and below it equal $1$. Then $\bigl[\widehat M{}^{-p}_{12}\bigr]_{i,j}=(-1)^{i-j}{{p}\choose{i-j}}$ and the numbers
\be
f^r_p:= \widehat M_{22}\bigl[\widehat M{}^{-1}_{12}\bigr]^p  \widehat M_{11}
\ee
are subject to the recursion $f^r_{p+1}=f^r_p-f^{r-1}_p$ with the initial data $f^r_1=f^1_p=1$ for all $p,r\ge 1$. The unique solution of this recursion is
\be
f^r_p=(-1)^{r-1} {{p-2}\choose{r-1}},
\ee
and in particular, $f^r_p=0$ for $1<p\le r$. Since $f^r_p$ tell us how many matrix products sum up (with signs) into $M_{22}\bigl[M_{12}^{-1}\bigr]^p M_{11}$, and it is easy to check that actual cancellations also occur at the level of products of $n\times n$ elementary transport matrices, we conclude that $M_{22}\bigl[M_{12}^{-1}\bigr]^p M_{11}$ are zeros for $1<p\le r$ and nonzero for $p>r$ for a generic choice of elementary transport matrices. Therefore, in this example, we have the loop algebra with shifted $T^-(u)$: \ $T^-_k:= M_{22}M_{12}^{-r-k}M_{11}$, $k\ge 1$.
\end{example}

\subsection{Twisted Yangians}\label{ss:Yang}
A standard procedure (see, e.g., Example 13 in chapter 2 of \cite{Molev}) produces a twisted Yangian satisfying the quantum reflection equation from the quantum loop algebra.

\begin{theorem}
Having the quantum loop algebra (see Definition~\ref{def:loop}) with the quantum $R$-matrix (\ref{R-affine}), the combinations
\be
\label{Au}
\mathbb A(u):= \bigl[ T^-(u^{-1}) \bigr]^{\text{T}} T^+(u) = \sum_{k=0}^\infty A^{(k)} u^{-k}
\ee
satisfy the quantum affine reflection equation
\be
\label{TRTR-u}
R(u,v) \sheet{1}{\mathbb A}(u) R^{t_1}(u^{-1},v) \sheet{2}{\mathbb A}(v)=\sheet{2}{\mathbb A}(v) R^{t_1}(u^{-1},v) \sheet{1}{\mathbb A}(u)R(u,v).
\ee
Moreover, the zero-level terms $ A^{(0)}=[T^-_1]^{\text{T}} T^+_0$ satisfy the quantum reflection equation
\be
\label{TRTR}
R \sheet{1}{A}{}^{(0)} R^{t_1} \sheet{2}{A}{}^{(0)}=\sheet{2}{A}{}^{(0)} R^{t_1} \sheet{1}{A}{}^{(0)} R.
\ee
\end{theorem}

The {\bf proof} is a short calculation, which we reproduce for the presentation integrity. We first find commutation relations for transposed matrices: total transposition of (\ref{R++}) or (\ref{R--}) gives (recall that the order of products in the quantum space is not affected by the transposition operation)
$$
\bigl[\sheet{1}T{}^\pm(u) \bigr]^{\text{T}} \bigl[\sheet{2}T{}^\pm(v) \bigr]^{\text{T}} [u R^{-1} -v R^{\text{T}}]= [u R^{-1} -v R^{\text{T}}]  \bigl[\sheet{2}T{}^\pm(v) \bigr]^{\text{T}}  \bigl[\sheet{1}T{}^\pm(u) \bigr]^{\text{T}}.
$$ 
Using now that $[u R -v R^{-\text{T}}] [u R^{-1} -v R^{\text{T}}]=(u^2+v^2-(q^2+q^{-2})uv)I$ and that $[u^{-1} R -v^{-1} R^{-\text{T}}]=\frac{-1}{uv} [u R^{-\text{T}}-v R]$, we finally obtain
\be
R(u,v) \bigl[\sheet{1}T{}^\pm(u^{-1}) \bigr]^{\text{T}} \bigl[\sheet{2}T{}^\pm(v^{-1}) \bigr]^{\text{T}} = \bigl[\sheet{2}T{}^\pm(v^{-1}) \bigr]^{\text{T}}  \bigl[\sheet{1}T{}^\pm(u^{-1}) \bigr]^{\text{T}} R(u,v)
\ee
Next, perform the transposition w.r.t. the first space in (\ref{R-+}) and replace the argument $u\to u^{-1}$:
$$
\bigl[\sheet{1}T{}^-(u^{-1}) \bigr]^{\text{T}} [u^{-1}(R^{-1})^{t_2} - v R^{t_1}] \sheet{2}T{}^+(v) = \sheet{2}T{}^+(v) [u^{-1}(R^{-1})^{t_2} - v R^{t_1}]  \bigl[\sheet{1}T{}^-(u^{-1}) \bigr]^{\text{T}}.
$$
It is easy to check that interchanging spaces $1\leftrightarrow 2$ and arguments $u \leftrightarrow v$ leaves this equation invariant:
$$
\bigl[\sheet{2}T{}^-(v^{-1}) \bigr]^{\text{T}} [u^{-1}(R^{-1})^{t_2} - v R^{t_1}] \sheet{1}T{}^+(u) = \sheet{1}T{}^+(u) [u^{-1}(R^{-1})^{t_2} - v R^{t_1}]  \bigl[\sheet{2}T{}^-(v^{-1}) \bigr]^{\text{T}}.
$$
We need one more relation on products of $R$-matrices: for all $u,v,w,y\in \mathbb C$, we have that
$$
R^{t_1}(u,v) R(w,y)= R(w,y) R^{t_1}(u,v). 
$$
Armed with all these commutation relations, we obtain
\begin{align*}
&R(u,v) \bigl[\sheet{1}T{}^-(u^{-1}) \bigr]^{\text{T}}  \sheet{1}T{}^+(u) R^{t_1}(u^{-1},v) \bigl[\sheet{2}T{}^-(v^{-1}) \bigr]^{\text{T}}  \sheet{2}T{}^+(v) \\
=&R(u,v) \bigl[\sheet{1}T{}^-(u^{-1}) \bigr]^{\text{T}} \bigl[\sheet{2}T{}^-(v^{-1}) \bigr]^{\text{T}}  R^{t_1}(u^{-1},v) \sheet{1}T{}^+(u)  \sheet{2}T{}^+(v) \\
=&\bigl[\sheet{2}T{}^-(v^{-1}) \bigr]^{\text{T}}  \bigl[\sheet{1}T{}^-(u^{-1}) \bigr]^{\text{T}}  R(u,v) R^{t_1}(u^{-1},v) \sheet{1}T{}^+(u)  \sheet{2}T{}^+(v)\\
=&\bigl[\sheet{2}T{}^-(v^{-1}) \bigr]^{\text{T}}  \bigl[\sheet{1}T{}^-(u^{-1}) \bigr]^{\text{T}}   R^{t_1}(u^{-1},v) R(u,v) \sheet{1}T{}^+(u)  \sheet{2}T{}^+(v)\\
=&\bigl[\sheet{2}T{}^-(v^{-1}) \bigr]^{\text{T}}  \bigl[\sheet{1}T{}^-(u^{-1}) \bigr]^{\text{T}}   R^{t_1}(u^{-1},v)   \sheet{2}T{}^+(v) \sheet{1}T{}^+(u) R(u,v) \\
=&\bigl[\sheet{2}T{}^-(v^{-1}) \bigr]^{\text{T}}    \sheet{2}T{}^+(v)  R^{t_1}(u^{-1},v)   \bigl[\sheet{1}T{}^-(u^{-1}) \bigr]^{\text{T}}  \sheet{1}T{}^+(u) R(u,v).
\end{align*}
So, the main statement of the theorem is proved. A similar calculation produce commutation relations for matrix entries of $A^{(0)}$.\quad $\square$


\section{Perspectives}
The goal of this paper was to demonstrate that for every quantum cluster variety (a {\em seed} in Fomin--Zelevinsky terminology \cite{FZ}) corresponding to a planar network on a disc with separated $m_1$ sources and $m_2$ sinks we can construct a finite (parameterized by $0<m<\min(m_1,m_2)$) family of leaves of Lie--Poisson quantum affine algebras (Theorem~\ref{th:A}). Moreover, if the $m\times m$ transport matrix $M_{12}$ is invertible in the direct product of classical and quantum spaces, then we can define the full quantum loop  algebra (Theorem~\ref{th:loop}) from which, by the standard procedure, we can construct a (finite-dimensional) symplectic leaf of the affine quantum reflection equation. We can therefore construct a {\em quantum cluster realization} for any such leaf and can use it to exploit possible representations of the corresponding affine algebras. Another important promising direction is to relate cluster structures appearing in the affine Lie--Poisson algebras to symplectic structures arising \cite{AlMal} from classic Lie algebra representations.  

All proofs in this paper used only $R$-matrix relations. Note that the quantum algebra of transport elements is well defined for {\em any} planar network, not necessarily with separated sources and sinks (see \cite{ChSh2}). A first natural problem is to generalize the method of this paper to the case of interlacing sources and sinks. Another observation pertains to the quantum ordering procedure: it is a Weyl ordering for all elements of the quantum transport matrices $M_{\alpha\beta}$, but in the products defining affine elements, the products of operator-valued entries of $M_{\alpha\beta}$ are taken in the order coinciding with the order of entries in the matrix product. Is it the only possible prescription for products of quantum entries? 

Note that, unlike the case in \cite{ChSh2} where the elements of quantum reflection equation defined for a standard $SL_n$ network in the disc with three complete flags (three marked points) constitute an upper-triangular matrix, the zero-level matrix ${A}{}^{(0)}$ is not upper triangular even for an $SL_n$ network on an annulus, so it is important to find out the conditions ensuring its canonical upper-triangular form. It is especially interesting in the affine case, because if the matrix ${A}{}^{(0)}$ is upper-triangular, we know~\cite{ChM} the affine action of the braid group, which is instrumental in studies of related invariants of knots and links. 

Another problem, also mentioned in \cite{ChM}, is related to reductions of the quantum loop algebras. In a realization where elements of the loop algebra correspond to powers of an (invertible) matrix $M_{12}$, reductions for which $A^{(k)}\equiv 0$ for $k>N$ are impossible; nevertheless, we can require a periodicity condition ensured by the restriction that $M_{12}^N=\mathbf I$ for some natural number $N$ (in \cite{ChM}, where elements of $\mathbb A$ were obtained by rotations about the central hole of the annulus, this condition was ensured by the requirement that this central hole was actually a $\mathbb Z_N$-orbifold point). These constraints are Lagrangian and result in finite-dimensional reductions of the quantum loop algebra, and one can hope that for a proper choice of $M_{12}$ we might be able to parameterize the higher dimensional symplectic leaves of such Poisson reductions of quantum loop algebras/quantum reflection equations in terms of cluster variables.

\section*{Acknowledgements}
The author is grateful to Michael and Alexander Shapiro for the useful discussion. The work was partially financially supported by RFBR Grant No. 18-01-00460.

\setcounter{section}{0}
\appendix{Commutation relations for $T^-_2$}
We calculate the commutation relations between two matrices $T^-_2:=M_{22}[M{}_{12}^{-1}]^2 M_{11}$. We also use the notation $T^-_3:=M_{22}[M{}_{12}^{-1}]^3 M_{11}$ and $T^-_1:=M_{22}M{}_{12}^{-1} M_{11}$.
\begin{align*}
&R^{-\text{T}}  \sheet{1}M_{22} \sheet{1}M{}_{12}^{-2}\tcb{ \sheet{1}M_{11} \sheet{2}M_{22}}  \sheet{2}M{}_{12}^{-2}  \sheet{2}M_{11}\\
=&R^{-\text{T}}  \sheet{1}M_{22} \sheet{1}M{}_{12}^{-2}\tcb{ \sheet{2}M_{22}}\tcr{  \sheet{1}M_{11} } \sheet{2}M{}_{12}^{-2}  \sheet{2}M_{11}+
R^{-\text{T}}  \sheet{1}M_{22} \sheet{1}M{}_{12}^{-1}\tcr{ \sheet{1}M{}_{12}^{-1}} (q{-}q^{-1})\sheet{2}M_{21} \tcr{ \sheet{1}M_{12} }P \sheet{2}M{}_{12}^{-2} 
\sheet{2}M_{11}\\
=&\tcb{ R^{-\text{T}}  \sheet{1}M_{22} \sheet{2}M_{22}} \bigl[R^{-1}\sheet{1}M{}_{12}^{-1}\bigr]^2 \tcr{\sheet{2}M{}_{12}^{-1}} R^{-\text{T}}  \sheet{2}M{}_{12}^{-1}  \tcb{  R^{-\text{T}}  \sheet{1}M_{11}  \sheet{2}M_{11}}\\
&\quad +R^{-\text{T}}  \sheet{1}M_{22} \sheet{1}M{}_{12}^{-1}\tcr{ \sheet{1}M{}_{12}^{-1}} (q{-}q^{-1})\sheet{2}M_{21} \tcr{ \sheet{1}M_{12} }P \sheet{2}M{}_{12}^{-2} \sheet{2}M_{11} \\
=&  \sheet{2}M_{22}  \tcr{ \sheet{1}M_{22}R^{-\text{T}} \sheet{2}M{}_{12}^{-1}}   \sheet{1}M{}_{12}^{-1}R^{-1}\sheet{1}M{}_{12}^{-1}\tcb{R^{-1} R^{-\text{T}} } \sheet{2}M{}_{12}^{-1}  \sheet{2}M_{11}  \sheet{1}M_{11}R^{-\text{T}} +
R^{-\text{T}}  \sheet{1}M_{22} (q{-}q^{-1})\sheet{2}M_{21} P \sheet{2}M{}_{12}^{-3} \sheet{2}M_{11} \\
=&  \sheet{2}M_{22} \sheet{2}M{}_{12}^{-1}  \sheet{1}M_{22}   \sheet{1}M{}_{12}^{-1}R^{-1}\sheet{1}M{}_{12}^{-1}\tcb{ [I- (q{-}q^{-1}) R^{-1}P]} \sheet{2}M{}_{12}^{-1}  \sheet{2}M_{11}  \sheet{1}M_{11}R^{-\text{T}} \\
&\quad +R^{-\text{T}}  \sheet{1}M_{22} (q{-}q^{-1})\sheet{2}M_{21} P \sheet{2}M{}_{12}^{-3} \sheet{2}M_{11} \\
=&  \sheet{2}M_{22} \sheet{2}M{}_{12}^{-1}  \sheet{1}M_{22}   \sheet{1}M{}_{12}^{-1}\tcb{R^{-1}\sheet{1}M{}_{12}^{-1} \sheet{2}M{}_{12}^{-1} } \sheet{2}M_{11}  \sheet{1}M_{11}R^{-\text{T}} \\
&\quad - (q{-}q^{-1}) \sheet{2}M_{22} \sheet{2}M{}_{12}^{-1}  \sheet{1}M_{22}   \sheet{1}M{}_{12}^{-1}R^{-1}\sheet{1}M{}_{12}^{-1}  R^{-1} \sheet{1}M{}_{12}^{-1}  \sheet{1}M_{11}  \tcr{\sheet{2}M_{11}R^{-1}}P\\
&\quad +(q{-}q^{-1}) R^{-\text{T}}  \sheet{1}M_{22} \sheet{2}M_{21} P \sheet{2}M{}_{12}^{-3} \sheet{2}M_{11} \\
=&  \sheet{2}M_{22} \sheet{2}M{}_{12}^{-1}  \sheet{1}M_{22}  \tcb{ \sheet{1}M{}_{12}^{-1} \sheet{2}M{}_{12}^{-1}} \tcr{ \sheet{1}M{}_{12}^{-1} R^{-1} \sheet{2}M_{11} } \sheet{1}M_{11}R^{-\text{T}} \\
&\quad - (q{-}q^{-1}) \sheet{2}M_{22} \sheet{2}M{}_{12}^{-1}  \tcr{\sheet{1}M_{22} \sheet{2}M_{11} }  \sheet{1}M{}_{12}^{-1}\sheet{1}M{}_{12}^{-1} \sheet{1}M{}_{12}^{-1}  \sheet{1}M_{11} P\\
&\quad +(q{-}q^{-1}) R^{-\text{T}}  \sheet{1}M_{22} \sheet{2}M_{21} P \sheet{2}M{}_{12}^{-3} \sheet{2}M_{11} \\
=&  \sheet{2}M_{22} \sheet{2}M{}_{12}^{-1}  \sheet{1}M_{22}  R \sheet{2}M{}_{12}^{-1} \tcr{ \sheet{1}M{}_{12}^{-1}   R^{-1} \sheet{2}M_{11} } \sheet{1}M{}_{12}^{-1} \sheet{1}M_{11}R^{-\text{T}} \\
&\quad + (q{-}q^{-1})^2 \sheet{2}M_{22} \tcr{\sheet{2}M{}_{12}^{-1}}  \sheet{1}M_{21} \tcr{\sheet{2}M_{12} } P \bigr[ \sheet{1}M{}_{12}^{-1}\bigl]^3 \sheet{1}M_{11} P\\
&\quad +(q{-}q^{-1}) R^{-\text{T}}  \sheet{1}M_{22} \sheet{2}M_{21} P \sheet{2}M{}_{12}^{-3} \sheet{2}M_{11}   - (q{-}q^{-1}) \sheet{2}T{}^-_1  \sheet{1}T{}^-_3P  \\
=&  \sheet{2}M_{22} \sheet{2}M{}_{12}^{-1}  \sheet{1}M_{22} \tcb{ R} \sheet{2}M{}_{12}^{-1} \sheet{2}M_{11}  \sheet{1}M{}_{12}^{-1}    \sheet{1}M{}_{12}^{-1} \sheet{1}M_{11}R^{-\text{T}} \\
&\quad + (q{-}q^{-1})\tcr{\bigl[ (q{-}q^{-1})P+ R^{-\text{T}}\bigr] }\sheet{1}M_{22}  \sheet{2}M_{21} \bigr[ \sheet{1}M{}_{12}^{-1}\bigl]^3 \sheet{1}M_{11} P
  - (q{-}q^{-1}) \sheet{2}T{}^-_1  \sheet{1}T{}^-_3P  \\
=&  \sheet{2}M_{22} \sheet{2}M{}_{12}^{-1}  \sheet{1}M_{22} \tcb{\bigl[ (q{-}q^{-1})P+ R^{-\text{T}}\bigr] } \sheet{2}M{}_{12}^{-1} \sheet{2}M_{11}  \sheet{1}M{}_{12}^{-1}    \sheet{1}M{}_{12}^{-1} \sheet{1}M_{11}R^{-\text{T}} \\
&\quad + (q{-}q^{-1})\tcr{R \sheet{1}M_{22}  \sheet{2}M_{21} }\bigr[ \sheet{1}M{}_{12}^{-1}\bigl]^3 \sheet{1}M_{11} P
  - (q{-}q^{-1}) \sheet{2}T{}^-_1  \sheet{1}T{}^-_3P  \\
=&  \sheet{2}M_{22} \sheet{2}M{}_{12}^{-1} \tcb{  \sheet{1}M_{22}R^{-\text{T}}  \sheet{2}M{}_{12}^{-1} } \sheet{2}M_{11}  \sheet{1}M{}_{12}^{-1}    \sheet{1}M{}_{12}^{-1} \sheet{1}M_{11}R^{-\text{T}} \\
&\quad + (q{-}q^{-1})P \sheet{1}M_{22} \sheet{1}M{}_{12}^{-1} \bigl( \sheet{2}M_{22}  \sheet{2}M{}_{12}^{-1} \sheet{2}M_{11}\bigr)  \sheet{1}M{}_{12}^{-1}    \sheet{1}M{}_{12}^{-1} \sheet{1}M_{11}R^{-\text{T}} \\
&\quad + (q{-}q^{-1})\tcr{R \sheet{1}M_{22}  \sheet{2}M_{21} }\bigr[ \sheet{1}M{}_{12}^{-1}\bigl]^3 \sheet{1}M_{11} P
  - (q{-}q^{-1}) \sheet{2}T{}^-_1  \sheet{1}T{}^-_3P  \\
=&  \sheet{2}M_{22} \sheet{2}M{}_{12}^{-1} \sheet{2}M{}_{12}^{-1} \tcb{  \sheet{1}M_{22}  \sheet{2}M_{11} } \sheet{1}M{}_{12}^{-1}    \sheet{1}M{}_{12}^{-1} \sheet{1}M_{11}R^{-\text{T}} \\
&\quad + (q{-}q^{-1})P \sheet{1}M_{22}\bigl[ \sheet{1}M{}_{12}^{-1} \bigr]^3 \tcb {\bigl( \sheet{2}M_{22}  \sheet{2}M{}_{12}^{-1} \sheet{2}M_{11}\bigr)  \sheet{1}M_{11}R^{-\text{T}} }\\
&\quad + (q{-}q^{-1}) \sheet{2}M_{21}  \sheet{1}M_{22}\bigl[ \sheet{1}M{}_{12}^{-1}\bigr]^3 \sheet{1}M_{11} P
  - (q{-}q^{-1}) \sheet{2}T{}^-_1  \sheet{1}T{}^-_3P  \\
=&  \sheet{2}M_{22} \sheet{2}M{}_{12}^{-1} \sheet{2}M{}_{12}^{-1}  \sheet{2}M_{11}   \sheet{1}M_{22}  \sheet{1}M{}_{12}^{-1}    \sheet{1}M{}_{12}^{-1} \sheet{1}M_{11}R^{-\text{T}} \\
&\quad -(q{-}q^{-1}) 
 \sheet{2}M_{22} \sheet{2}M{}_{12}^{-1} \sheet{2}M{}_{12}^{-1}  \sheet{1}M_{21}  \sheet{2}M_{12} P  \sheet{1}M{}_{12}^{-1}    \sheet{1}M{}_{12}^{-1} \sheet{1}M_{11}R^{-\text{T}} \\
&\quad + (q{-}q^{-1})P \sheet{1}M_{22}\bigl[ \sheet{1}M{}_{12}^{-1} \bigr]^3  \sheet{1}M_{11} \sheet{2}M_{22}  \sheet{2}M{}_{12}^{-1} \sheet{2}M_{11}\\
&\quad - (q{-}q^{-1})^2P \sheet{1}M_{22}\bigl[ \sheet{1}M{}_{12}^{-1} \bigr]^3  \sheet{2}M_{21}   \sheet{1}M_{11}P\\
&\quad - (q{-}q^{-1})\bigl[ \sheet{2}T{}^-_1- \sheet{2}M_{21} \bigr]  \sheet{1}T{}^-_3P  \\
=&  \sheet{2}T{}^-_{2} \sheet{1}T{}^-_{2}R^{-\text{T}}  -(q{-}q^{-1})P
 \sheet{1}M_{22}\bigl[ \sheet{1}M{}_{12}^{-1}\bigr]^3 \sheet{2}M_{21} \sheet{1}M_{11}\tcb{ \bigl[R^{-\text{T}}+(q{-}q^{-1})P\bigr] }\\
&\quad + (q{-}q^{-1})P\sheet{1}T{}^-_{3} \sheet{2}T{}^-_{1}  - (q{-}q^{-1})\bigl[ \sheet{2}T{}^-_1- \sheet{2}M_{21} \bigr]  \sheet{1}T{}^-_3P  \\
=&  \sheet{2}T{}^-_{2} \sheet{1}T{}^-_{2}R^{-\text{T}}  -(q{-}q^{-1})P
 \sheet{1}M_{22}\bigl[ \sheet{1}M{}_{12}^{-1}\bigr]^3 \tcb{ \sheet{2}M_{21} \sheet{1}M_{11} R }\\
&\quad + (q{-}q^{-1})P\sheet{1}T{}^-_{3} \sheet{2}T{}^-_{1}  - (q{-}q^{-1})\bigl[ \sheet{2}T{}^-_1- \sheet{2}M_{21} \bigr]  \sheet{1}T{}^-_3P  \\
=&  \sheet{2}T{}^-_{2} \sheet{1}T{}^-_{2}R^{-\text{T}}  -(q{-}q^{-1})P  \sheet{1}T{}^-_{3}  \sheet{2}M_{21} 
 + (q{-}q^{-1})P\sheet{1}T{}^-_{3} \sheet{2}T{}^-_{1}  - (q{-}q^{-1})\bigl[ \sheet{2}T{}^-_1- \sheet{2}M_{21} \bigr]  \sheet{1}T{}^-_3P  \\
=&  \sheet{2}T{}^-_{2} \sheet{1}T{}^-_{2}R^{-\text{T}}  + (q{-}q^{-1})P\sheet{1}T{}^-_{3}\bigl[ \sheet{2}T{}^-_{1}- \sheet{2}M_{21}\bigr]  - (q{-}q^{-1})\bigl[ \sheet{2}T{}^-_1- \sheet{2}M_{21} \bigr]  \sheet{1}T{}^-_3P  .
  \end{align*}
 Note that if the groupoid condition $M_{22}M_{12}^{-1}M_{11}-M_{21}=0$ is satisfied, we obtain homogeneous commutation relations $R^{-\text{T}}\sheet{1}T{}^-_{2} \sheet{2}T{}^-_{2} =\sheet{2}T{}^-_{2} \sheet{1}T{}^-_{2}R^{-\text{T}} $.

\end{document}